\documentclass[apj]{emulateapj} 
\usepackage{apjfonts} 

\shortauthors{White et al.}
\shorttitle{Stacked {\it FIRST} Quasars}

\slugcomment{To be published in ApJ}

\newcommand{\muJy}{\mbox{$\mu$Jy}}
\newcommand{\mJy}{\mbox{mJy}}
\newcommand{\GHz}{\mbox{GHz}}
\newcommand{\MUV}{\mbox{$M_{UV}$}}
\newcommand{\LR}{\mbox{$L_R$}}
\newcommand{\LRM}{\mbox{$L_{R,M}$}}
\newcommand{\RM}{\mbox{$R^*_M$}}
\newcommand{\prim}{{\it primary\/}}
\newcommand{\hiz}{{\it high-z\/}}

\begin{document}
\title{Signals from the Noise: Image Stacking for Quasars in the FIRST Survey}
\author{
Richard L.~White\altaffilmark{1},
David J.~Helfand\altaffilmark{2},
Robert H.~Becker\altaffilmark{3,4},
Eilat Glikman\altaffilmark{2,5},
Wim deVries\altaffilmark{3,4}}
\email{rlw@stsci.edu}

\altaffiltext{1}{Space Telescope Science Institute, Baltimore, MD 21218}
\altaffiltext{2}{Dept. of Astronomy, Columbia University, New York, NY 10027}
\altaffiltext{3}{Physics Dept., University of California, Davis, CA 95616}
\altaffiltext{4}{IGPP/Lawrence Livermore National Laboratory}
\altaffiltext{5}{Dept. of Astronomy, Caltech, Pasadena, CA 91125}

\begin{abstract}

We present a technique to explore the radio sky into the nanoJansky
regime by employing image stacking using the {\it FIRST}
survey.  We begin with a discussion of the non-intuitive relationship
between the mean and median values of a non-Gaussian distribution
in which measurements of the members of the distribution are dominated
by noise. Following a detailed examination of the systematic effects
present in the 20~cm VLA snapshot images that comprise {\it FIRST},
we demonstrate that image stacking allows us to recover the average
properties of source populations with flux densities a factor of
30 or more below the rms noise level.  With the calibration described
herein, mean estimates of radio flux density, luminosity, radio
loudness, etc. are derivable for any undetected source class having
arcsecond positional accuracy.

We demonstrate the utility of this technique by exploring the radio
properties of quasars found in the Sloan Digital Sky Survey. We
compute the mean luminosities and radio-loudness parameters for
41,295 quasars in the SDSS DR3 catalog.
There is a tight
correlation between optical and radio luminosity, with the radio
luminosity increasing as the 0.85 power of optical luminosity.
This implies declining radio-loudness with optical luminosity:
the most luminous objects ($\MUV = -28.5$) have on average three
times lower radio-to-optical ratios than the least luminous objects
($\MUV = -20$).  There is also a striking correlation between
optical color and radio loudness: quasars that are either redder or
bluer than the norm are brighter radio sources.  Quasars having
$g-r \sim 0.8$ magnitudes redder than the SDSS composite spectrum
are found to have radio-loudness ratios that are higher by
a factor of 10. We explore the longstanding question of whether a
radio-loud/radio-quiet dichotomy exists in quasars, finding that
optical selection effects probably dominate the distribution function
of radio loudness, which has at most a modest ($\sim 20$\%) inflection
between the radio-loud and radio-quiet ends of the distribution.
We also examine the radio properties of the
subsample of quasars with broad absorption lines,
finding, surprisingly, that BAL quasars have {\it higher} mean radio
flux densities at all redshifts, with the greatest disparity arising
in the rare low-ionization BAL subclass. We conclude with examples of
other problems for which the stacking analysis developed here is
likely to be of use.

\end{abstract}


\section{Introduction}

``Blank sky'' is rarely truly blank. All astronomical imaging observations
have a sensitivity threshold below which ``objects'' are not
detectable. Assuming a reasonably linear detector response, however, it
is not necessarily the case that zero photons from discrete sources have
been detected at a given ``blank'' spot in an image. If one has reason
to believe from observations in another wavelength regime that discrete
emitters are present at a set of well-specified locations, it is possible 
to usefully constrain, or even to detect, the mean flux of that set of
emitters by stacking their ``blank sky'' locations.
The prerequisites for successfully stacking images
in this way are good astrometry for both the target objects and
the survey images, and sufficient sky coverage to include a
large sample of the target class.

In an early application
of this technique, Caillault \& Helfand (1985) detected the mean X-ray
flux from undetected G-stars in the Pleiades, using
it to constrain the decay of stellar X-ray emission with age. As 
higher-resolution X-ray mirrors and detectors have become available over the past two
decades, X-ray stacking has become a standard analysis technique. Applications
have ranged from determining the mean X-ray luminosity of object
classes in deep X-ray images --- e.g., normal galaxies (Brandt et al.\ 2001a),
Lyman Break galaxies (Brandt et al.\ 2001b), and radio sources (Georgakakis et 
al. 2003) --- to determining the X-ray cluster emission from distant
clusters in the Rosat All-Sky Survey (Bartelmann \& White 2003).

As linear digital detectors have come to dominate optical and infrared sky
surveys, the stacking technique has been widely adopted: e.g., Zibetti et al.\
(2005) detected intracluster light by stacking 683 Sloan Digital
Sky Survey (SDSS\footnote{Funding
for the SDSS and SDSS-II has been provided by the Alfred P. Sloan
Foundation, the Participating Institutions, the National Science
Foundation, the U.S. Department of Energy, the National Aeronautics
and Space Administration, the Japanese Monbukagakusho, the Max
Planck Society, and the Higher Education Funding Council for England.
The SDSS Web Site is \url{http://www.sdss.org}.})
clusters, Lin et al.\ (2004) stacked 2MASS data on cluster galaxies,
Hogg et al.\ (1997) stacked
Keck IR data to get faint galaxy colors, and Minchin et al.\ (2003) went so far 
as to stack digitized films from the UK Schmidt telescope to complement a deep
\ion{H}{1} survey with the Parkes multi-beam receiver. Scaramella et al.\ (1993) have
even stacked cosmological simulations in investigating the Sunyaev-Zeldovich
effect on the cosmic microwave background.

The radio sky is relatively sparsely populated with sources. The
deepest large-area sky survey, {\it FIRST}, has a surface density
of only $\sim 90$ sources deg$^{-2}$ at its 20~cm flux density
threshold of 1.0~mJy. Fluctuation analysis of the deepest radio images
ever made suggest a source surface density of $\sim 15$ arcmin$^{-2}$
at $\sim 1~\muJy$ (Windhorst et al.\ 1993); given that the mean
angular size of such sources is $\sim 2.4\arcsec$, even at these
flux density levels only $\sim3\%$ of the sky is covered by radio
emission. Nonetheless, applications of stacking in the radio band
have been limited. Recently, Serjeant at el. (2004) stacked SCUBA data
to find the mean submillimeter flux of Spitzer $24~\mu$m-selected
galaxies. It is with large-area surveys and large counterpart
catalogs, however, that the stacking technique allows one to reach extremely faint
flux density levels unachievable by direct observations.

The {\it FIRST} survey (Becker, White, \& Helfand 1995) is ideally
suited for stacking studies. It contains 811,000 sources brighter
than $\sim1~\mJy$ over 9030 deg$^{2}$ of the northern sky and has
an angular resolution of $5\arcsec$. Thus, over 99.9\% of its five
billion beam areas represent blank sky. Having written several dozen
papers on sources detected in the survey, we turn here to analyzing
the remaining 99.9\% of the data. Glikman et al.\ (2004b) presented
our initial results of applying radio image stacking to the {\it
FIRST} survey. Wals et al.\ (2005) applied this technique to the 2dF
quasar catalog (Croom et al.\ 2004) using {\it FIRST} images, producing
an estimate of the mean flux density for undetected quasars in the
range $20-40~\muJy$. In this paper, we describe the set of detailed
tests we have carried out to calibrate biases in the {\it FIRST} survey's
VLA images in order to put stacking results on a quantitative basis,
and we illustrate the technique with several examples. Subsequent
papers will apply these results to various problems of interest.

We begin (\S2) with a discussion of the median stacking procedure
that we have adopted, exploring in some detail the meaning of, and
distinctions between, mean and median values in data dominated by noise.
We go on to provide a thorough analysis
of the noise characteristics of the {\sl FIRST} images, both by
stacking known sub-threshold sources and by the use of artificial
sources inserted into the survey data (\S3). We find a non-linear correction
to the flux densities derived from a stacking analysis, which most
likely arises from the application of the highly nonlinear `CLEAN'
algorithm to these undersampled $uv$ (snapshot) data. We then
apply our calibrated stacking procedure to the SDSS DR3
quasar survey from Schneider et al.\ (2005) ({\S4}).
In addition to deriving quasar radio properties
as a function of redshift and optical color, we reexamine
the issue of whether the radio-loudness distribution is
bimodal.
We also explore the distinction
between broad absorption line (BAL) and non-BAL objects, finding the surprising result that
BAL quasars have a higher mean flux density and radio loudness than
non-BAL objects below 2~mJy. We conclude ({\S5}) with a summary of
the implications of our results, and preview other applications of
our stacking procedure.

\section{Mean Versus Median Stacking Procedures}

We have explored two different methods for stacking sub-threshold
{\it FIRST} images, one using the mean of each pixel in the stack
and the other using the median.  Both approaches have advantages and disadvantages.
The mean flux density in a stacked image is mathematically simple
and is easily interpretable. However, it is very sensitive to the
rare outliers in the distribution.  The presence of a bright source
in the stack, either at the image center or in the periphery, makes
itself obvious in the summed image.  Moreover, noise outliers can
also cause problems, as a minority of very noisy images may
substantially raise the noise in the mean image.  The outlier problem
can be addressed by testing each image in the stack, discarding
sources that are actually above the {\it FIRST} detection threshold and/or discarding 
images that exceed some rms noise threshold.  However, the resulting mean is
sensitive to the exact value of the discard threshholds and hence
does not provide a very robust measurement.

The alternative is to determine the median value of the stacked
images. The obvious advantage is the insensitivity of the median
to outliers, since the median is well known to be robust in the
presence of non-Gaussian distributions (e.g., Gott et al.\ 2001).
Therefore all of the data can be utilized, eliminating the need to
impose an arbitrary cutoff to the distribution. However, the
interpretation of the median value for low signal-to-noise (S/N)
data is not straightforward. For high S/N data, the median is simply
the value at the midpoint of the distribution.  But in the case of
low S/N data, the value obtained by taking the median is shifted
from the true median toward the `local' mean value.
The degree of the shift depends on the amplitude of the noise; as
the noise increases, the recovered value approaches the `local'
mean, where the `local' mean is the mean of the values within
approximately one rms of the median.  Hence the recovered median
value can depend on both the intrinsic distribution of the parameter
and the noise level.
\begin{figure*}
\epsscale{0.95}
\plottwo{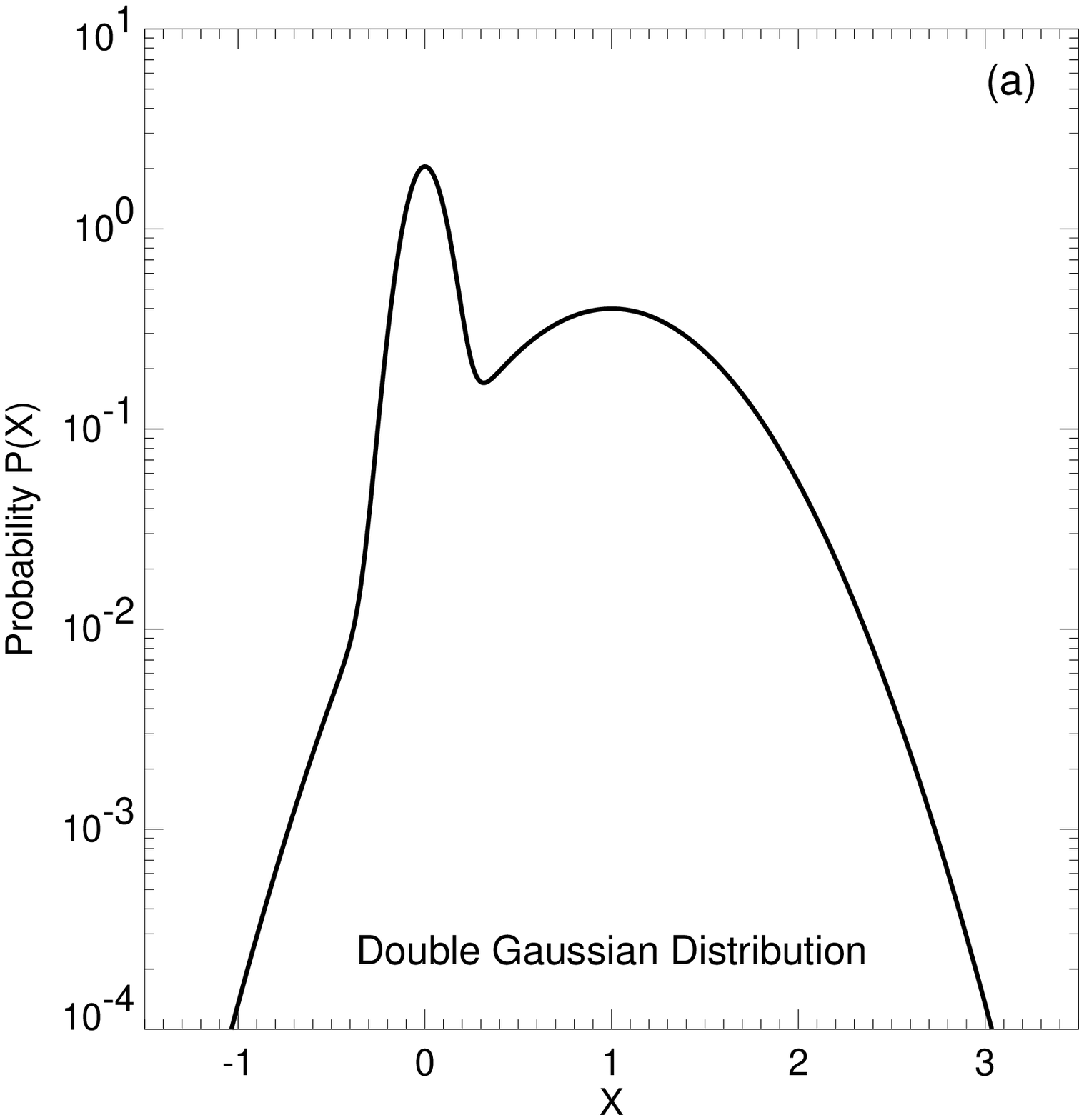}{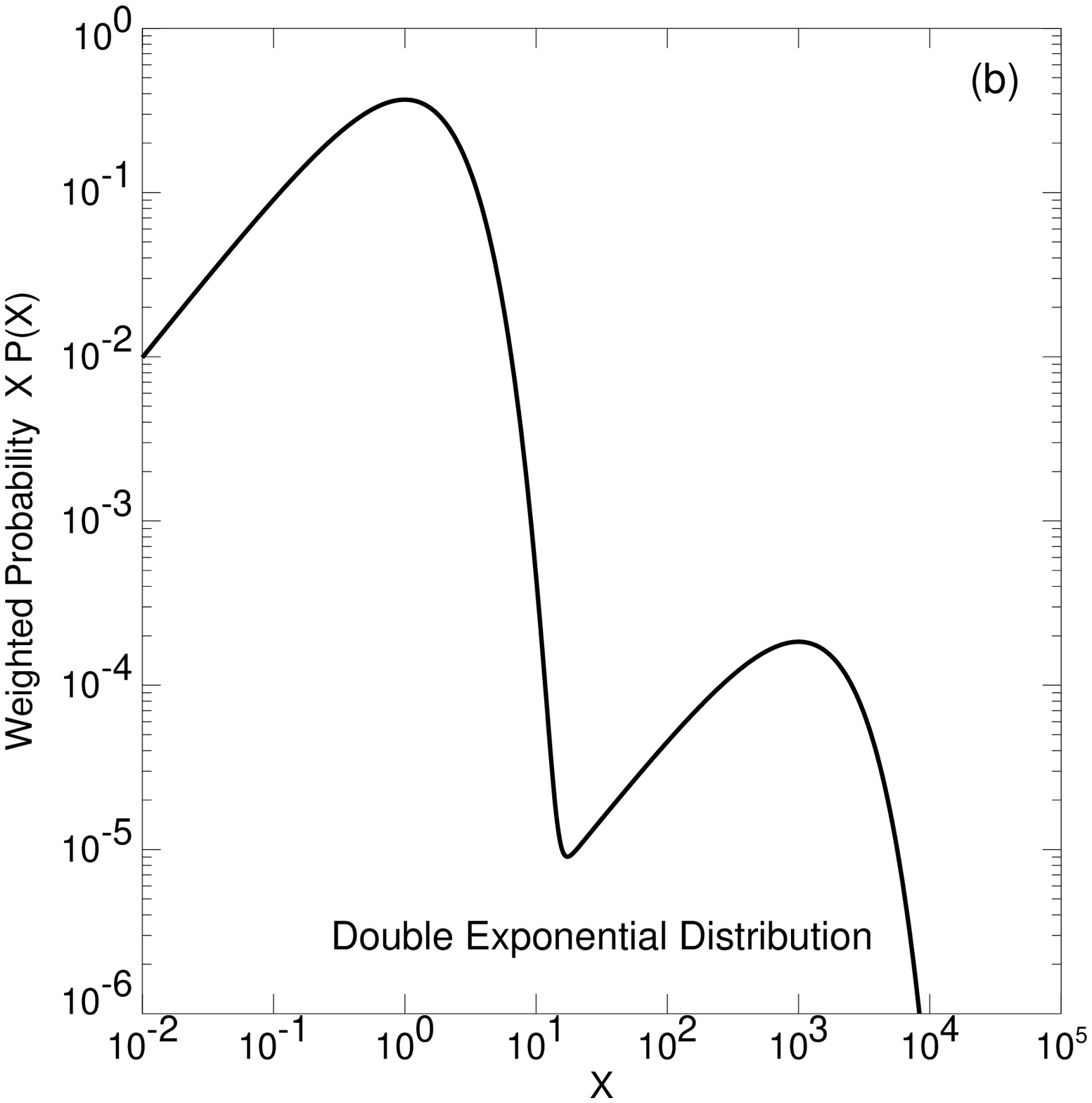}
\plottwo{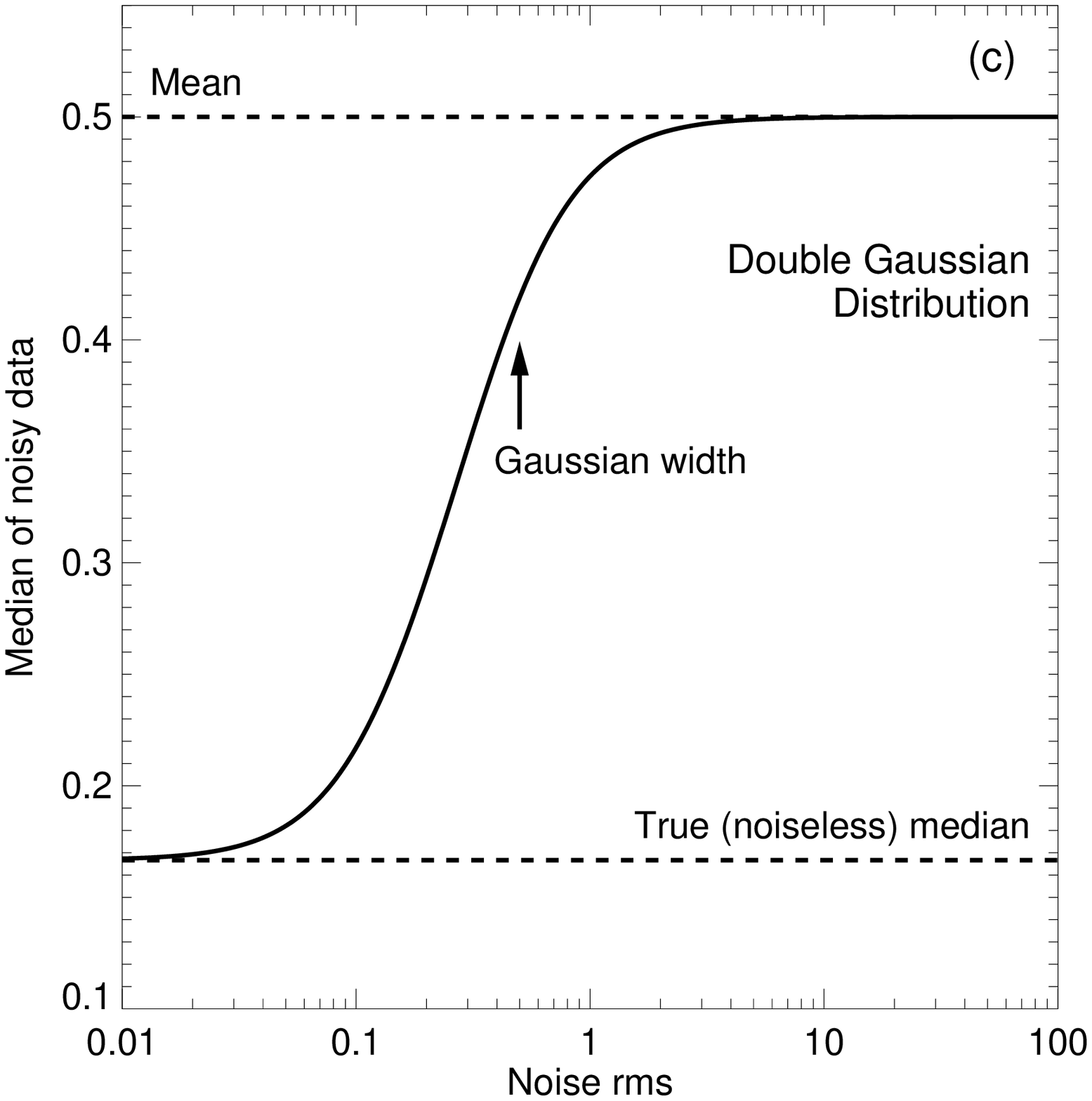}{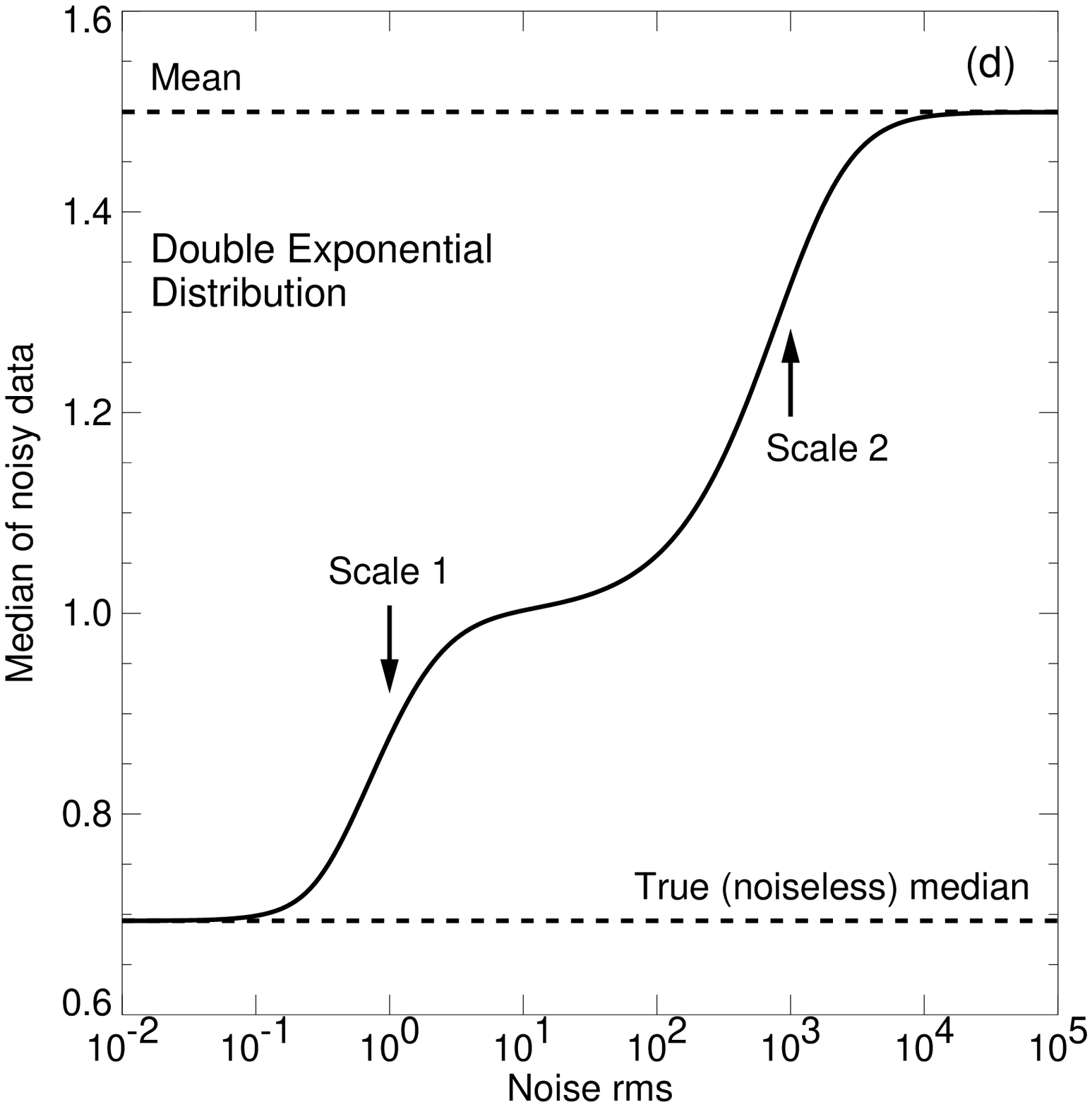}
\epsscale{1.0}
\caption{The effect of measurement noise on the measured median
value for a skewed distribution. (a) A distribution consisting of
two equal Gaussians, a wide component at $x=1$ and a narrow component
at $x=0$. (b) A distribution composed of two exponentials having
very different scale heights with the broad component including
only a small fraction of the population.  The probability has been
multiplied by $x$ for a better visual display of the
distribution plotted versus $\log\,x$.  (c) The median for the double Gaussian
distribution with noise added shows a smooth transition.  When the
noise is small, we recover the true median, and as the rms noise
becomes comparable to the separation of the components, the value
converges to the mean for the distribution.  (d) The median for
the double exponential distribution with noise presents a transition
from the true median to the mean that pauses at the mean for the
dominant (narrower) population.
}
\label{fig-median_vs_mean}
\end{figure*}
Some concrete examples may help to illuminate this effect.
Figure~\ref{fig-median_vs_mean} shows the dependence of the measured
median value (computed using numerical integration)
on the noise level for two asymmetrical distributions.
First consider a simple distribution consisting of two Gaussians
centered at $x_1=0$ and $x_2=1$ with widths $\sigma_1=0.1$ and
$\sigma_2=0.5$ (Fig.~\ref{fig-median_vs_mean}a).  The Gaussians are
normalized to have equal integral amplitudes, but because the first
is much narrower, its peak is higher by a factor of $\sigma_2/\sigma_1
= 5$.  The mean of the distribution is midway between the Gaussians
at $\langle x\rangle=0.5$, but the median is dominated by the much
better-localized narrow component and falls at median$(x)=0.167$.
If samples are drawn from this distribution with additive Gaussian
measurement noise, the peaks of both distributions get broadened.  In
the limit where the noise is much larger than $x_2-x_1$, the median
value converges to the mean $\langle x\rangle$.  The transition
with increasing values of the noise rms is shown in
Figure~\ref{fig-median_vs_mean}(c).

The general shape of the transition in Figure~\ref{fig-median_vs_mean}(c)
is typical for simple skewed distributions (e.g., power laws,
exponentials, etc.).  However, more complicated distributions display
a more complex dependence on the noise level.
Figure~\ref{fig-median_vs_mean}(b) shows 
a distribution composed of two one-sided exponentials, $P(x)dx
= \exp(-x/h)/h$, with $x>0$.  The first exponential drops rapidly,
with a scale height $h_1=1$, while the second drops much more slowly,
$h_2=1000$.  The two components are normalized so that the first
contains the vast majority of the sources, with the integrated
amplitude of the second component being only 0.05\% of the total.
When Gaussian noise is added, there are three separate regimes of
behavior (Fig.~\ref{fig-median_vs_mean}d).  When the rms noise is
much smaller than either exponential scale, the true median is
recovered: median$(x)=0.694$, just slightly above the median
computed for component 1 only ($\ln2 = 0.693$).  For very large rms
noise levels the measured median converges to the mean for the whole
distribution ($\langle x\rangle = 1.50$).  But for intermediate
values of the rms around unity, there is an inflection where the
measured median value pauses at a value of median$(x)\sim 1$.
This is explained by the fact that the dominant exponential
with $h_1=1$ is itself a skewed distribution having a mean $\langle
x\rangle_1 = 1$.

It is worth noting that the standard arithmetic mean is also of limited utility
in the presence of complex multiple-component, strong-tailed
distributions like that in Figure~\ref{fig-median_vs_mean}(b).  Even
for those distributions the median is generally a better match to
one's intuitive concept of the ``typical'' value of the distribution.

Despite these complications (about which we have found little
discussion in the astronomical literature), we believe that the
median is distinctly preferable to the mean for stacking our {\it FIRST}
survey images.  In our tests, the robust median calculation produces
significantly more stable results with lower noise, while giving
very similar measured values for the fluxes.  We are in a limit
where almost all the values in our sample are small compared with
the noise, so it is straightforward to interpret our median stack
measurements as representative of the mean for the population of
sources with flux densities fainter than a few times the {\it FIRST} rms
(i.e., a few $\times0.145~\mJy$). Throughout the paper, we refer to
the median-derived approximate mean interchangeably as the ``median'' or
``average'' of the quantity of interest.

\section{Calibration of the Stacking Procedure}

\subsection{Introduction}

As an aperture synthesis interferometer, the Very Large Array\footnote{The
Very Large Array is an instrument of the National Radio Astronomy
Observatory, a facility of the National Science Foundation operated
under cooperative agreement by Associated Universities, Inc.} samples
the Fourier transform of the radio brightness distribution on the
sky. To obtain a sky image requires transforming to the image plane
with incomplete information. The 165-second snapshots that comprise
the {\it FIRST} survey are particularly problematic in this regard:
we typically obtain $\sim 30,000$ visibility points and transform
them into images with $\sim 5\times10^5$ resolution elements. The
nonlinear algorithm `CLEAN' (H{\"o}gbom 1974; Clark 1980) is used to
minimize artifacts such as the diffraction spikes produced by the
VLA geometry and the grating rings imposed by the minimum antenna
spacings employed.

One consequence of this process, discovered in the course of the
NVSS (Condon et al.\ 1998) and {\sl FIRST} surveys, is ``CLEAN
bias''. This incompletely understood phenomenon steals flux from
the above-threshold sources and redistributes it around the field.
The magnitude of the bias is dependent on the rms noise in the image
(it increases as noise increases), the off-axis angle (it decreases
in consort with the primary beam pattern), and the source extent
(extended sources lose more flux). The {\sl FIRST} and NVSS surveys
took considerable pains to calibrate CLEAN bias, concluding,
respectively, that it had values of 0.25~mJy/beam and 0.30~mJy/beam.
(It is unsurprising that the different resolutions, integration
times, and analysis procedures of the two surveys produced slightly
different results).

While the sources of interest in a stacking analysis are sub-threshold
and therefore, by definition, have not been CLEANed, we have taken
the discovery of CLEAN bias as a cautionary tale and have examined
in detail the behavior of our images subjected to the stacking
process. We find that we do not recover the full flux density of
either artificial sources inserted into the images or real sub-threshold
sources. We describe here our calibration of this phenomenon, which
we dub ``snapshot bias''.

\subsection{Artificial source tests}

The Astronomical Image Processing System (AIPS) used to reduce VLA
data includes a task UVMOD that allows the user to insert artificial
sources into a $uv$ database in order to test the fidelity of the
analysis. Since in this case we are interested in sources far below
the detection threshold, a large number of artificial sources is
required. The median rms in coadded {\sl FIRST} fields is 145~\muJy; thus,
to achieve an uncertainty of $\sim 10\%$ in the measured flux
density of, say, 40~\muJy\ sources requires the addition of more than
1200 individual sources.

For our initial attempt to measure the bias, we inserted one hundred
40~\muJy\ sources placed in a regular square grid into each of 100
{\sl FIRST} fields.  This approach allowed us to minimize the number
of maps we needed to make.  However, this failed as a consequence
of the interference of the sidelobe patterns that even these very
faint sources produce.  Constructive and destructive interference of
sidelobes led to noise in the stacked images that varied 
strongly and systematically at different artificial source
grid positions.  We concluded that this test was not sufficient to
measure the bias for random source locations. 

Our ultimate artificial source test involved placing four 40~\muJy\
sources in each of 400 {\sl FIRST} $uv$ datasets. The sources were
placed at the corners of a square of side $\sim 2\arcmin$ centered on
the image.  We then CLEANed the 400 images, extracted $1\arcmin$
cutouts around each of the fake source locations, and stacked the
cutouts to find the median flux density.  Since artificial source
locations were not screened in advance, they ocassionally fell on
or near the location of a real radio source; the median algorithm
effectively rejected the contaminated pixels in those cases (\S2).
Source parameters were derived by fitting an elliptical Gaussian
to the stacked image as is done for source extraction in the real
images.  To improve the quality of the fits, regions around the
diffraction spikes in the VLA dirty beam (see Fig.~\ref{fig-dr3_stack}
below for an example) are masked out.  The process was repeated
for artificial sources with a peak flux density of 80~\muJy.

The results are presented in Table 1. The recovered median peak
flux densities for the 40~\muJy\ and 80~\muJy\ sources were
33~\muJy\ and 60~\muJy, respectively. The persistence of missing
flux reminiscent of the CLEAN bias at flux densities far below
those that experience CLEANing is a surprise; as shown below,
however, this result is confirmed by stacking results on faint
radio sources derived from deep, full-synthesis images.
\begin{deluxetable}{cccc}
\tablecolumns{4}
\tablecaption{Flux Density Bias in Stacked FIRST Images}
\tablehead{
\colhead{True Flux\tablenotemark{a}} & \colhead{Stack Median\tablenotemark{b}} &
\colhead{Bias\tablenotemark{c}} & \colhead{No. Images\tablenotemark{d}} \\
\colhead{($\mu$Jy)} & \colhead{($\mu$Jy)} &
\colhead{($\mu$Jy)} & \colhead{}
}
\startdata   
\cutinhead{Artificial Inserted Sources}
\phn\phn40 & \phn35 & \phn5 $\pm$ 3 & 1600 \\
\phn\phn80 & \phn61 &    19 $\pm$ 4 & 1600 \\
\cutinhead{First-Look Survey}
\phn182 & \phn134 & \phn48 $\pm$ 15 & 144 \\
\phn198 & \phn144 & \phn54 $\pm$ 14 & 145 \\
\phn219 & \phn154 & \phn65 $\pm$ 12 & 144 \\
\phn243 & \phn140 &    104 $\pm$ 15 & 145 \\
\phn277 & \phn203 & \phn74 $\pm$ 10 & 144 \\
\phn320 & \phn231 & \phn89 $\pm$ 27 & 145 \\
\phn385 & \phn288 & \phn98 $\pm$ 11 & 144 \\
\phn492 & \phn313 &    180 $\pm$ 14 & 145 \\
\phn733 & \phn503 &    230 $\pm$ 22 & 144 \\
1300\tablenotemark{e} & 1047 & 253 $\pm$ 41 & 145 \\
\cutinhead{COSMOS Survey}
\phn200 & \phn192 & \phn\phn8 $\pm$ 20 & 37 \\
\phn328 & \phn234 & \phn93 $\pm$ 61 & 38 \\
\phn594 & \phn354 &    240 $\pm$ 12 & 37 \\
1086\tablenotemark{e} & \phn893 & \phn193 $\pm$ 133 & 38 \\
\enddata   
\tablenotetext{a}{Mean peak flux density for sources in flux bin.}
\tablenotetext{b}{Median peak flux density for FIRST image stack.}
\tablenotetext{c}{Snapshot bias (underestimate of true flux) and rms uncertainty.}
\tablenotetext{d}{Number of sources and images in this bin.}
\tablenotetext{e}{This value is the median instead of the mean
because the noise in the individual images is small compared with
the bin's flux density range.}
\end{deluxetable}
\begin{figure}
\epsscale{1.15} 
\plotone{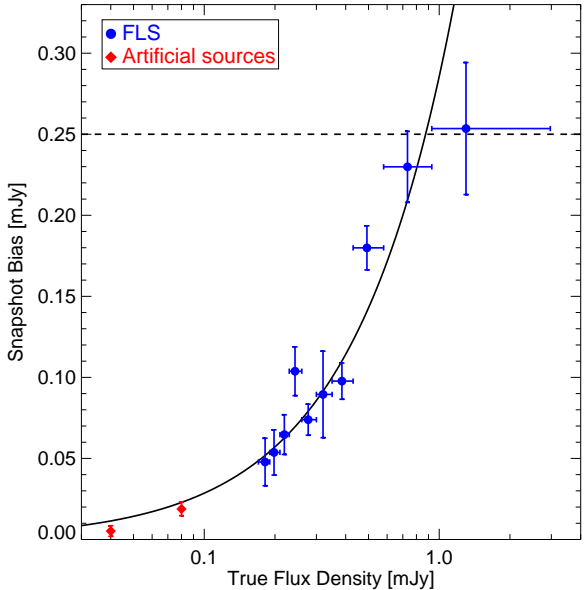}
\epsscale{1.0}
\caption{Snapshot bias for stacked FIRST images as a function of
the true flux density.  The bias is the difference between the true
flux density and the flux density measured in the median image.
The sample is limited to point sources (FWHM $<2.5\arcsec$), and
fitted peak values are used for the flux densities.  Data are shown for
both the The First-Look Survey (Condon et al.\ 2003; circles) and
artificial 40 and 80~\muJy\ sources (diamonds).  The vertical bars
indicate $1\sigma$ errors on the stacked flux density, and the horizontal
bars represent the range of flux densities for each bin.  The solid
line is a linear model in which the bias is 29\% of the true flux
density.  The dashed line is the 0.25~mJy CLEAN bias measured for sources
bright enough to be in the FIRST catalog.
}
\label{fig-deepbias}
\end{figure}

\subsection{Recovering real sub-threshold sources}

An alternative to using artificial sources is to stack real radio
sources detected in very deep VLA surveys.
The First-Look Survey (FLS -- Condon et al.\ 2003) covered the 5
deg$^2$ of the Spitzer First-Look fields using the VLA B~configuration;
it achieved a mean rms of 23~\muJy\ beam$^{-1}$ and detected 3565
sources down to a flux density of 115~\muJy. We ran our {\sl FIRST}
survey source extraction routine HAPPY (White et al.\ 1997) on the
publicly available FLS radio images and constructed a catalog of
1445 point-like sources (deconvolved source size $<2.5\arcsec$ with
the $5.0\arcsec$ beam) with flux densities ranging from 0.17 to
3.0~mJy; we chose a higher ($7\sigma$) source detection threshold to
minimize the uncertainties on the individual source flux densities.
We grouped the sources into ten equally populated flux density bins
with mean\footnote{The mean was used for the sub-threshold sources, since
this is the value to which our median stacking converges (see \S2), but the
median was used for the final bin, which contains detected {\it FIRST}
sources.} FLS fluxes ranging from 182~\muJy\ to 1300~\muJy.  We then
extracted $1\arcmin$ cutouts around each of these sources in the
{\sl FIRST} images and compared the true mean flux density (from
our FLS catalog) with the median stacked flux density in each bin.
Source parameters were derived by fitting an elliptical Gaussian
to the stacked image as described above.

We performed a similar analysis using data from the COSMOS
survey's pilot program (Schinnerer et al.\ 2004), comprised of
seven VLA pointings in the A~configuration that reached rms values
ranging from 36 to 46~\muJy.  The results are
consistent with the FLS survey, but the uncertainties are much
larger because the COSMOS sample has only one tenth as many sources as
the FLS sample.


The results are displayed in Table~1 and plotted in
Figure~\ref{fig-deepbias}.  For sources brighter than 0.75~mJy,
the mean deficit is consistent with the 0.25~mJy CLEAN bias we have
added to all above-threshold {\sl FIRST} sources (see above).
Below 0.75~mJy, however, the flux deficit changes character, and is
well-represented by a constant {\it fractional} offset, with the
stacked image yielding a value 71\% that of the true mean flux
density for each bin.

It is perhaps unsurprising that the bias should change near the
0.75~mJy threshold that divides brighter sources that were CLEANed
during the {\sl FIRST} image processing from fainter sources that
have not been CLEANed.  The continuity between the bias for
sub-threshold sources and that for super-threshold sources suggests
that they are aspects of the same phenonmenon.  A connection between
the bias and the depth of CLEANing is well established for
brighter sources (Becker, White \& Helfand 1995; Condon et al.\
1998), so we think it likely (but by no means certain) that the
snapshot bias is also created by the non-linear CLEAN process.

However, we are clueless as to why the
relationship has the particular form we observe.  According to
Cornwell, Braun \& Briggs (1999), ``to date no one has succeeded
in producing a noise analysis of CLEAN itself'', so we are not alone
in being mystified.  While we cannot offer a theoretical explanation
for snapshot bias, we will use the simple empirical bias correction:
\begin{equation}
\label{eqn-bias}
S_{p,corr} = \hbox{min}\left( 1.40\,S_p\, , \, S_p+0.25\,\mJy \right) \quad,
\end{equation}
where $S_p$ is the fitted peak flux density measured from the median stack.

Although Eq.~(\ref{eqn-bias}) was derived from elliptical Gaussian
fits to the stacked images, we find it applies equally well if the
brightness of the central pixel in the median image is used to
estimate the peak flux density.  We use both approaches below in the
analysis of the quasar sample.

Note that it is most fortunate that the bias is a constant fraction
of the flux, since that means that it can be corrected in the stacked
image.  That would not be true if, for example, it were a quadratic
function of flux, since in that case the bias in the summed image
would depend on the detailed distribution of contributing fluxes
(which is unknown). But since all faint sources have the same bias
correction multiplier, the bias correction can be appropriately
applied to the stacked image instead of the individual images.
In fact, it can be applied pixel-by-pixel to the stacked image
by simply multiplying each pixel in the image by 1.40.

\subsection{Stacking White Dwarfs, a Radio-Silent Population\label{section-wd}}

In the remainder of the paper we discuss the results of stacking
quasars divided into groups using many different parameters (redshift,
optical luminosity, etc.)  In all of the quasar subpopulations we
stack, we always detect a positive signal. In order to allay
concern that our algorithm somehow guarantees a detection, we have
stacked 2,412 white dwarfs from the SDSS DR1 white dwarf catalog
(Kleinman et al.\ 2004).  As expected, the stacked image shows no
hint of any source.  The image rms is $3.6~\muJy$, comparable to
the value expected from the typical FIRST rms of $145~\muJy$ divided
by $\sqrt N=49$.

\section{The Radio Properties of Undetected Quasars}

\begin{figure}
\epsscale{1.15} 
\plotone{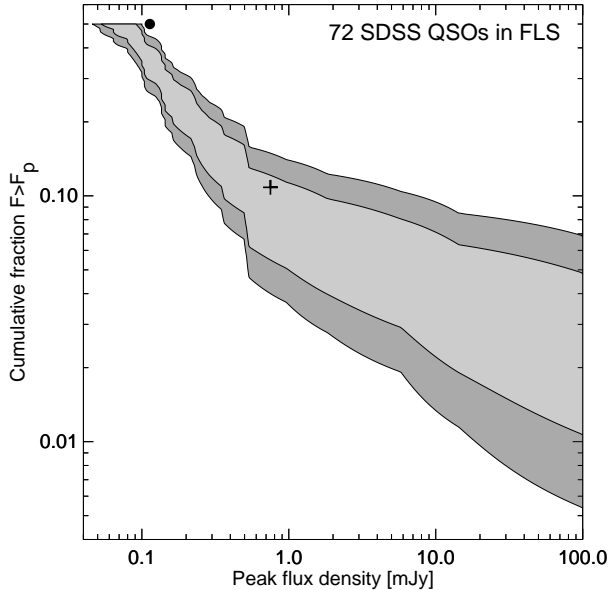}
\epsscale{1.0}
\caption{The cumulative fraction of radio-detected quasars as a
function of 20~cm flux density. The shaded bands represent the $\pm
1\sigma$ and $\pm90\%$ uncertainties derived using the FLS images
for the 72 SDSS quasars in the FLS survey area.  The cross at
0.75~mJy represents the fraction of all SDSS quasars detected in
the {\it FIRST} survey above this flux density, while the dot at
50\% fraction indicates the value of the median flux density derived
from our stacking analysis.  The
general agreement of the latter with the fraction of directly
detected quasars at these flux densities offers validation of our
approach, but the noise (which is comparable to the symbol size)
is far smaller in the values derived from
the stacked images.
}
\label{fig-fls_frac}
\end{figure}
Although radio emission was the defining feature of the first
quasars, more than four decades of effort has failed to establish
predictive models for quasar radio properties. While $\sim 10\%$
of quasars are relatively bright at centimeter wavelengths ($>
1~\mJy$) and thus are readily detected, the radio emission from
most quasars falls well below the limits of all large-area radio
surveys. Even deeper surveys that cover several square degrees of
sky only detect $\sim 50\%$ of quasars. For example, by examining
the images from the FLS radio survey described above (Condon et al.\ 2003),
we detect 36 of 72 SDSS quasars to a limiting flux density of $\sim
0.09$~mJy.  Figure~\ref{fig-fls_frac} shows the fraction of detected
sources as a function of flux density.  The width of the shaded
bands represent the $1\sigma$ and 90\% confidence uncertainties;
it is apparent that even this, the largest of all radio surveys to
this depth, is inadequate for determining accurately the detected
fraction as a function of flux density, let alone for understanding
how radio emission depends on redshift, absolute magnitude, the
presence of broad absorption lines, etc. For the forseeable future,
90\% of quasars will remain undetected at radio wavelengths. By
using image stacking with the {\it FIRST} survey, however, one can
begin to quantify the statistical properties of quasar radio emission
at all flux density levels.

\subsection{The radio properties of SDSS quasars}
\begin{figure*}
\epsscale{0.9}
\plotone{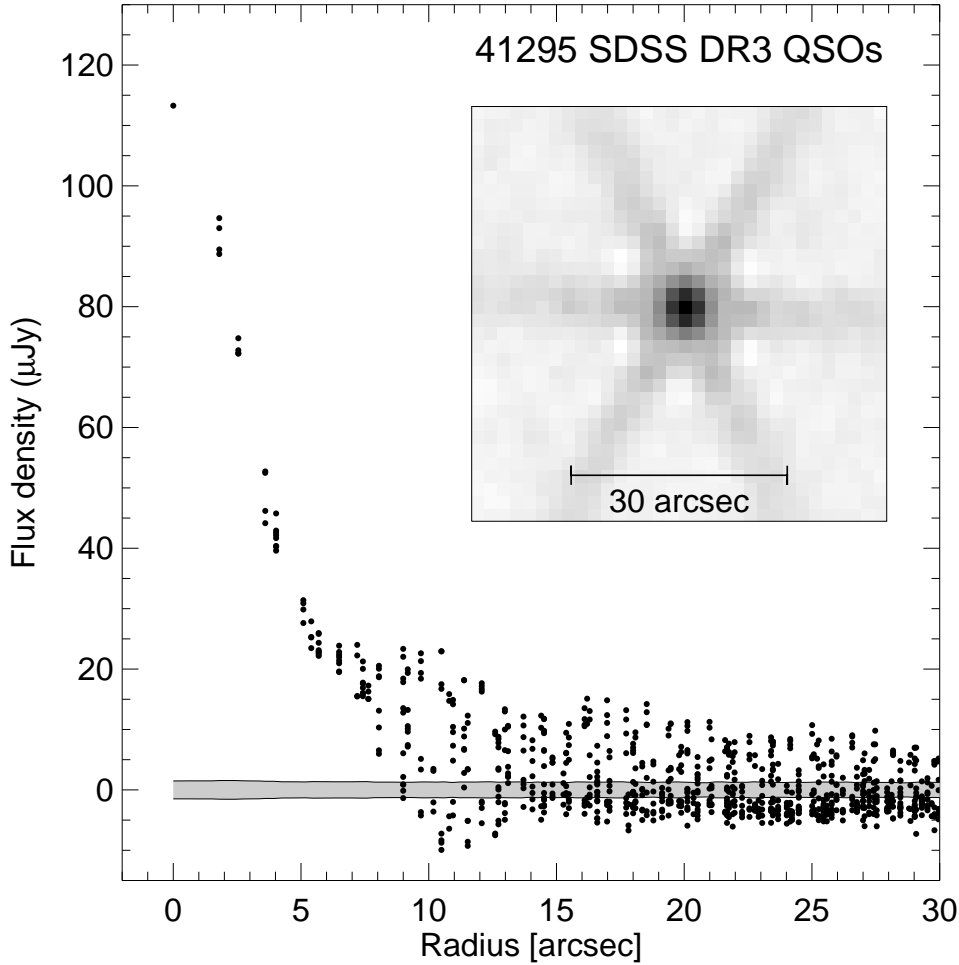}
\epsscale{1.0}
\caption{The result of constructing a median stack of the 41,295
source positions in the SDSS DR3 quasar catalog. The inset displays
a $1\arcmin$-square greyscale image (pixel size $1.8\arcsec$); the
positive and negative sidelobes of the VLA dirty beam pattern are
apparent.  The pixel-by-pixel radial plot shows the ``source''
profile with a FWHM of $\sim 7.0\arcsec$,
which is slightly extended compared with the PSF FWHM of $\sim5.4\arcsec$.
Flux density values 
have been corrected for snapshot bias (eqn.~\ref{eqn-bias}). The gray shaded band
indicates the $\pm 1\sigma$ errors calculated for median statistics
(Gott et al.\ 2001).}
\label{fig-dr3_stack}
\end{figure*}
As a starting point we use the largest existing quasar survey as
reported in the SDSS DR3 catalog (Schneider et al.\ 2005), which
contains 46,420 spectroscopically identified quasars. Of these
41,295 fall in regions covered by the {\it FIRST} survey.  Constructing
a median stack of the entire sample yields the image shown as the
inset in Figure~\ref{fig-dr3_stack}.  This high signal-to-noise
($\sim 75:1$) image shows a compact source centered on the nominal
quasar(s') position; a two-dimensional Gaussian fit yields a peak
raw flux density of $80~\muJy$, or roughly 50\% of the rms of an
individual {\it FIRST} image.  Multiplying this by 1.40 to correct
for the snapshot bias (Eq.~1) gives a peak flux density of $112\pm1.5~\muJy$.
The fluxes plotted in Fig.~\ref{fig-dr3_stack} and all fluxes quoted
hereafter have been corrected for the bias.

The six positive-flux radial spokes and interspersed negative
features are characteristic of the VLA sidelobe pattern; since the
vast majority ($\sim 93\%$) of sources contributing to this image
are below the {\it FIRST} detection threshold and are therefore not
CLEANed, this sidelobe pattern is expected.  Note that nearly all
FIRST fields are observed near the meridian so that sidelobes from
different fields align well.  The pixel-by-pixel radial profile in
Figure~\ref{fig-dr3_stack} shows a FWHM of $\sim7.0\arcsec$, slightly
larger than the size expected for a point source observed
in the VLA B~configuration at 20~cm.\footnote{The {\it FIRST}
survey's cataloged sources have a FWHM of $5.4\arcsec$ as a consequence
of the fact that the CLEANed {\it FIRST} images are convolved with
a CLEAN beam with that value; this is typically slightly larger
that the dirty beam size to accomodate images observed away from
the zenith where the synthesized beam shape is larger than the
nominal B-configuration value.} The shaded horizontal band indicates
the $\pm 1\sigma$ values derived for median statistics (Gott et
al.\ 2001).
\begin{figure}
\plotone{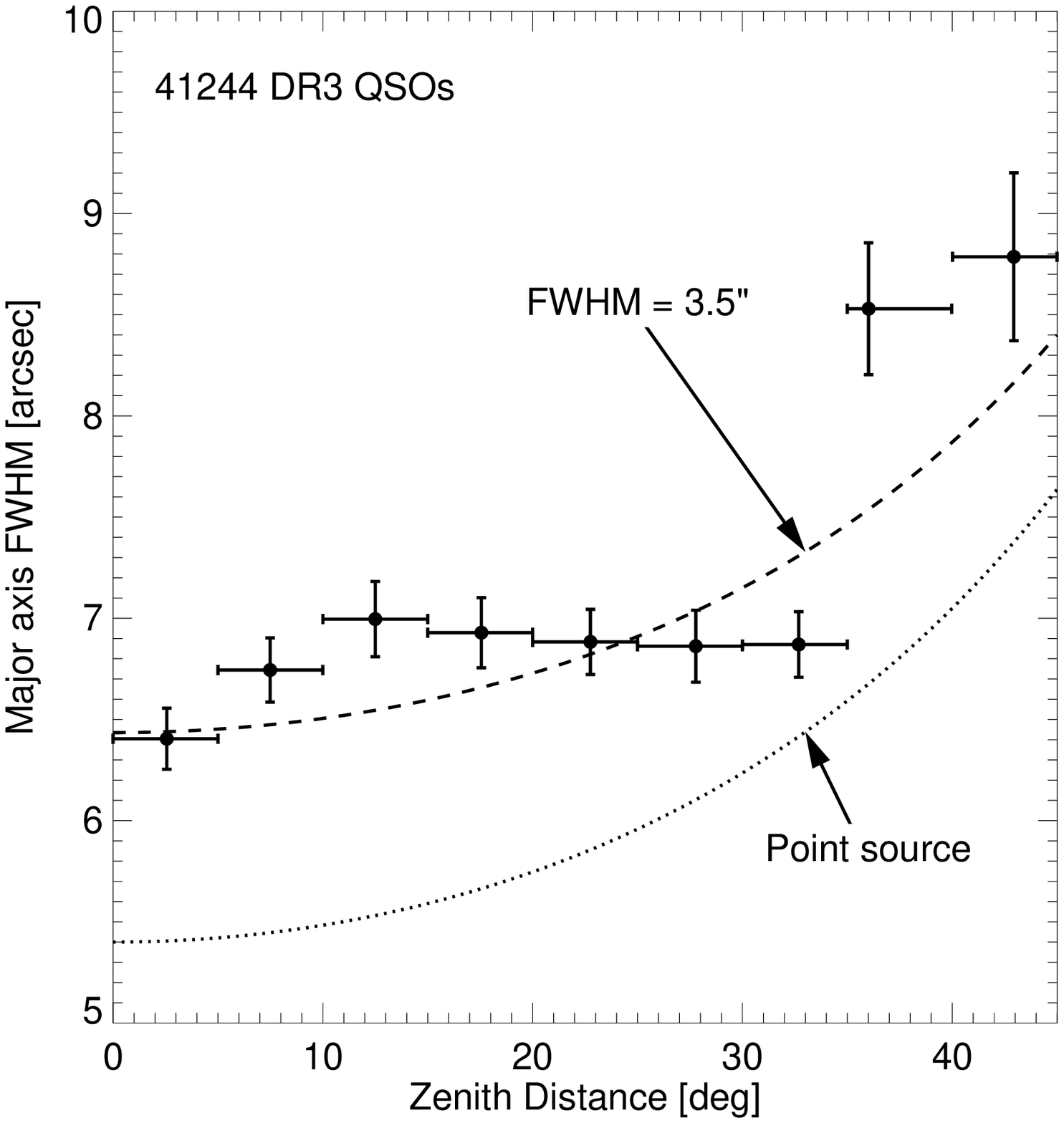}
\caption{
Size of the major axis for the stacked quasar images as a function
of the distance from the zenith at the VLA.  The VLA PSF width
increases with zenith distance due to foreshortening of the array
in the north-south direction.  The lines show the expected relationship
for a point source (dotted) and for a Gaussian source having a FWHM
of $3.5\arcsec$ (dashed).
}
\label{fig-size}
\end{figure}
The intrinsic radio source size implied by the extended emission
is affected by the VLA PSF size, which depends on the distance
of the source from the zenith.  Since most FIRST fields were observed
close to the meridian, the zenith distance is a simple function
of the source declination.  Figure~\ref{fig-size} shows the measured
radio sizes in image stacks separated into nine zenith-distance
bins.  The increase in size at high zenith distances is explained
by the increase in the VLA beam size.  Since the many quasars being
averaged for this measurement are randomly oriented, the stacked
radio image is expected to be symmetrical, and asymmetries are
explained by beam effects.  Both the distribution with
zenith distance and the fitted size for the image in
Figure~\ref{fig-dr3_stack} ($6\farcs4\times7\farcs0$ FWHM) are
consistent with a symmetrical quasar image having a mean source
size of 3\farcs5.  This is a bit larger than the size of
quasars at the 1~mJy detection limit of the FIRST survey.
Fitting the mean stack for the 679 quasars with central
flux densities between 1 and 2~mJy yields a size of $5\farcs8\times6\farcs2$,
implying an underlying source size of $2\farcs0\times3\farcs0$
when the beam size is deconvolved.

The median flux density of $\sim110~\muJy$ is reasonably consistent
with that found for the directly detected quasars within the FLS
sample (Fig.~\ref{fig-fls_frac}), though it is slightly higher than
the value in the FLS field ($74\pm20~\muJy$).  It is likely that
this difference is mainly the result of sample variance in
the FLS field.  If we stack the FIRST images for only the quasars
in the FLS fields, the flux density is $76\pm26~\mJy$, 
in good agreement with the measurement from the FLS images.

\subsection{Variation of radio properties with optical luminosity\label{section-lum}}
\begin{figure}
\plotone{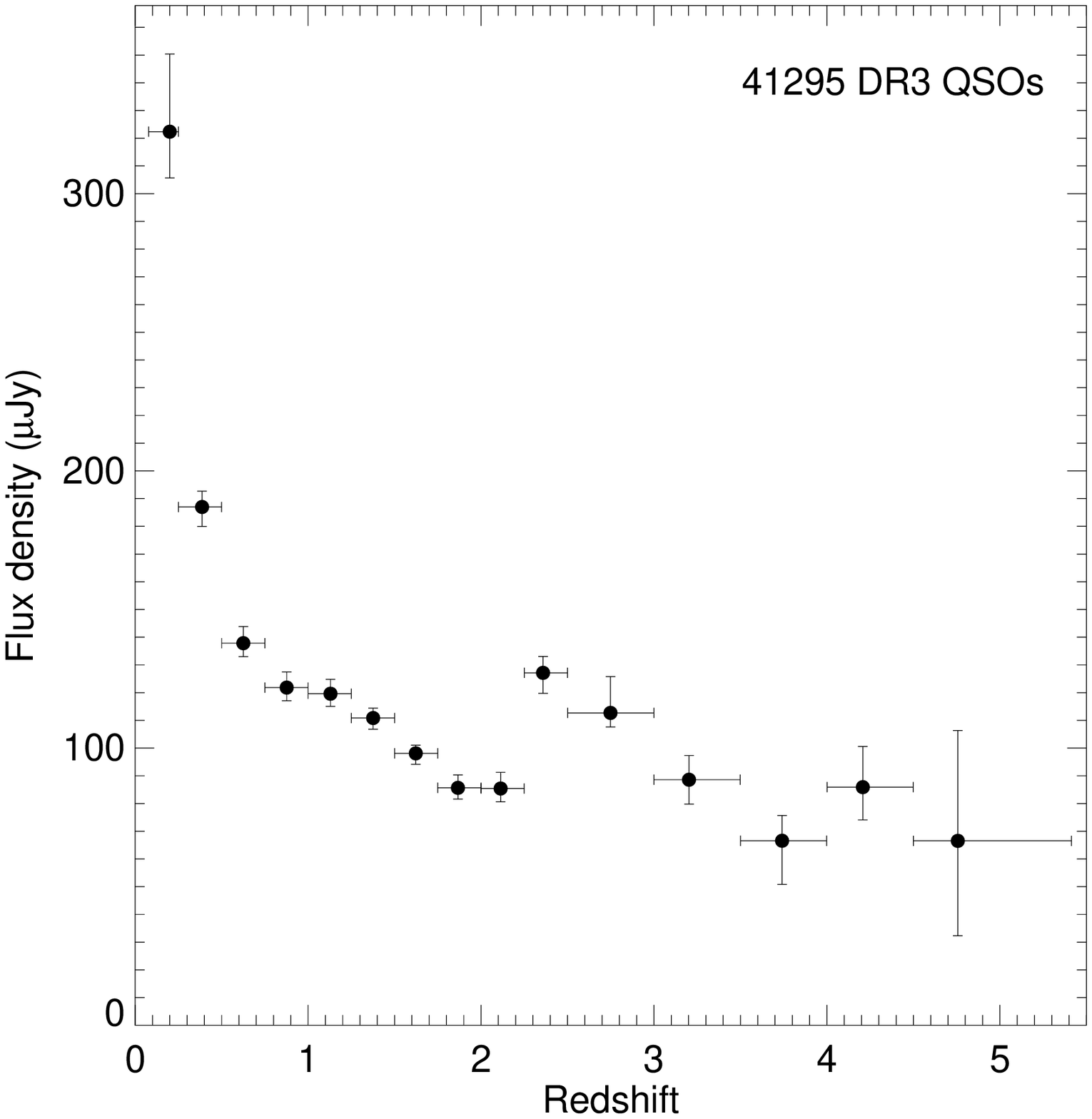}
\caption{The median flux density for SDSS DR3 quasars as a function of
redshift.}
\label{fig-flux_vs_z}
\end{figure}

\begin{figure}
\plotone{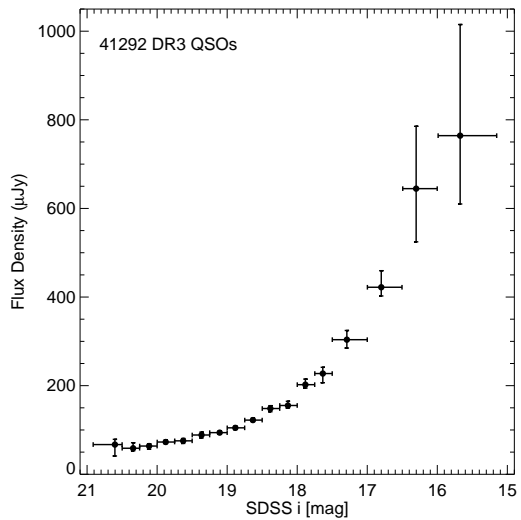}
\caption{
The median radio flux depends strongly on the SDSS $i$ magnitude.
Note that the for the brightest quasars the median flux approaches
the FIRST detection limit at 1~mJy.  This is consistent with the
conclusion from the FIRST Bright Quasar Survey that FIRST detects
most $V\sim15$ quasars (White et al.\ 2000.)
}
\label{fig-f_vs_i}
\end{figure}

\begin{figure}
\plotone{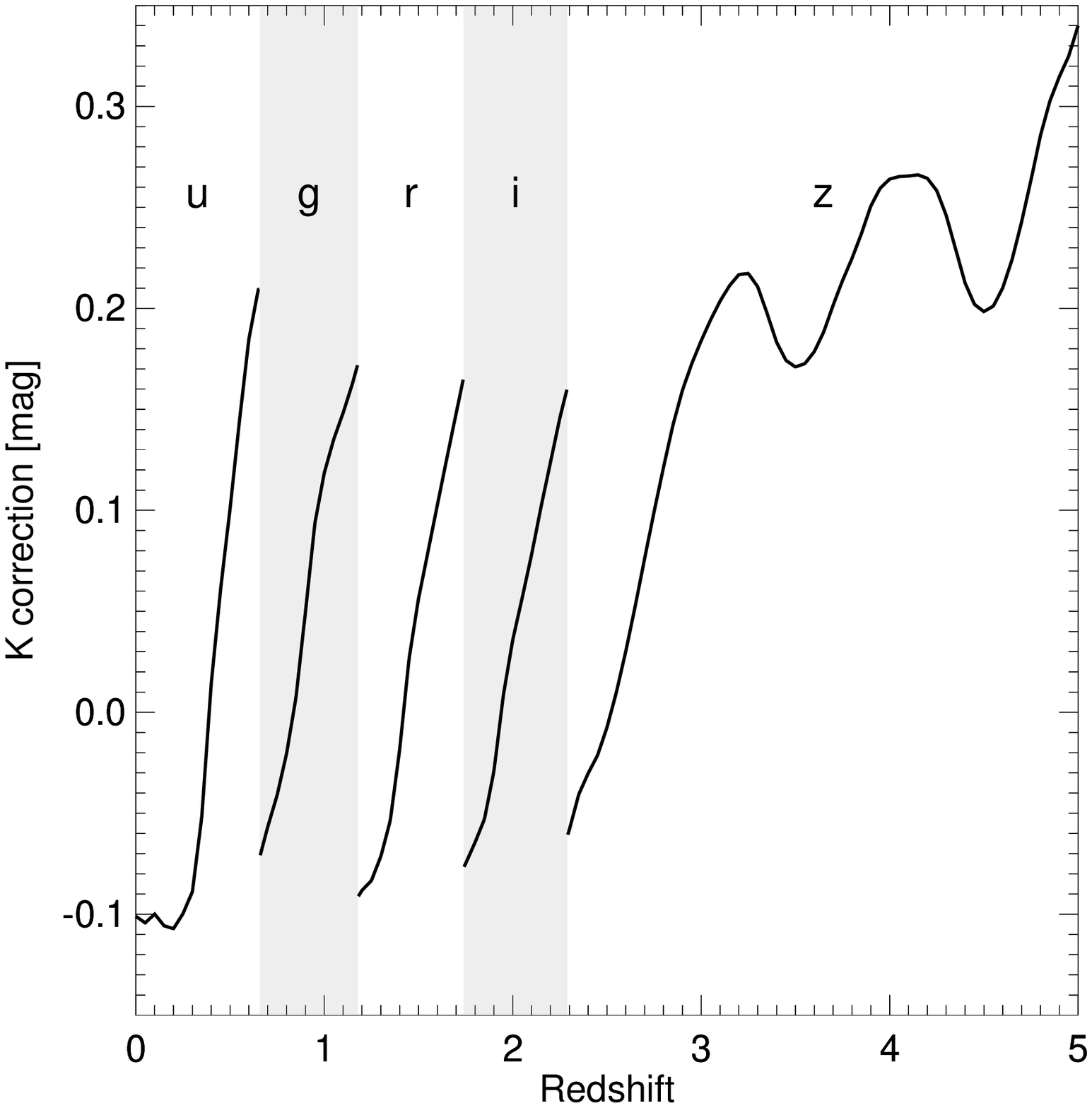}
\caption{
$K$-correction as a function of redshift to convert observed SDSS
magnitudes to the magnitude at 2500~\AA\ rest wavelength, derived
using the SDSS composite quasar spectrum (Vanden Berk et al.\ 2001).
The $K$-correction is added to the magnitude in the SDSS filter
closest to 2500~\AA\ (shown by the vertical bands).
The redshift-dependent bandwidth-stretching factor, $-2.5\log(1+z)$,
has been omitted here to make the plot clearer.
}
\label{fig-kcorr}
\end{figure}
Dividing the SDSS quasars into ten redshift bins, we see that the
median flux density declines monotonically to $z=2$
(Fig.~\ref{fig-flux_vs_z}).  At $z=2.25$ there is a noticeable jump
in the radio flux, which is a consequence of a confluence of effects
driven by the sharply declining efficiency of the SDSS quasar
selection algorithm (because the colors of $z=2$--3 quasars are
similar to stars; Richards et al.\ 2001, 2002) combined with an
interesting dependence of the radio emission on optical color
(discussed further below in \S\ref{section-redshift_color}).

The decline of flux with redshift is slower than the expected
scaling as the inverse of the luminosity-distance squared because
the SDSS sample is flux-limited and so detects increasingly luminous
objects as the redshift increases.  An interesting point is that
although the FIRST catalog is also flux-limited, the stacked FIRST
data are {\it not} flux-limited.  All sources get included in the
stack regardless of their radio brightnesses.  Consequently these
data do not suffer from the usual bias against faint sources in
the radio; only the optical flux limit introduces such a bias.
The radio luminosities are biased toward brighter values at high
redshifts only insofar as the radio and optical luminosities are
correlated.

The correlation between radio and optical luminosities does introduce
complications in interpreting our results.  Figure~\ref{fig-f_vs_i}
displays the radio flux as a function of SDSS $i$-band magnitude.
Optically bright sources are far more likely to be radio bright;
in fact, for the brightest quasars with $i<16$, the median radio
flux density approaches the 1~mJy detection limit for the FIRST
survey.  This is consistent with the conclusion from the FIRST
Bright Quasar Survey that FIRST detects most $V\sim15$ quasars
(White et al.\ 2000.)  But the potential entanglement of redshift,
absolute magnitude, and evolution makes it difficult to understand
the physical implications of this correlation.

We have concluded that the best approach is to correct for the
correlation between absolute magnitude and radio luminosity before
attempting to understand the variation in radio brightness with
secondary parameters.  We compute the 2500~\AA\ absolute magnitude,
\MUV, by applying a redshift-dependent $K$-correction derived using
the Vanden Berk et al.\ (2001) composite SDSS quasar spectrum
(Fig.~\ref{fig-kcorr}).  The $K$-correction is applied to the filter
closest to 2500~\AA\ at the quasar redshift.  The observed 20~cm
flux densities are converted to a rest-frame 5~GHz (6~cm) radio
luminosity, $\LR(5\,\GHz)$, using a spectral index of $\alpha =
-0.5$.  The redshift is converted to luminosity distance using a
standard WMAP cosmology ($\Omega_m=0.3$, $\Lambda=0.7$, $h=0.7$).
The particular choice of rest frame wavelengths facilitates comparison
of our results with previous studies using the radio-loudness
parameter $R^*$, defined by Stocke et al.\ (1992) as the ratio of
the 2500~\AA\ and 5~GHz flux densities.
\begin{figure}
\epsscale{1.15} 
\plotone{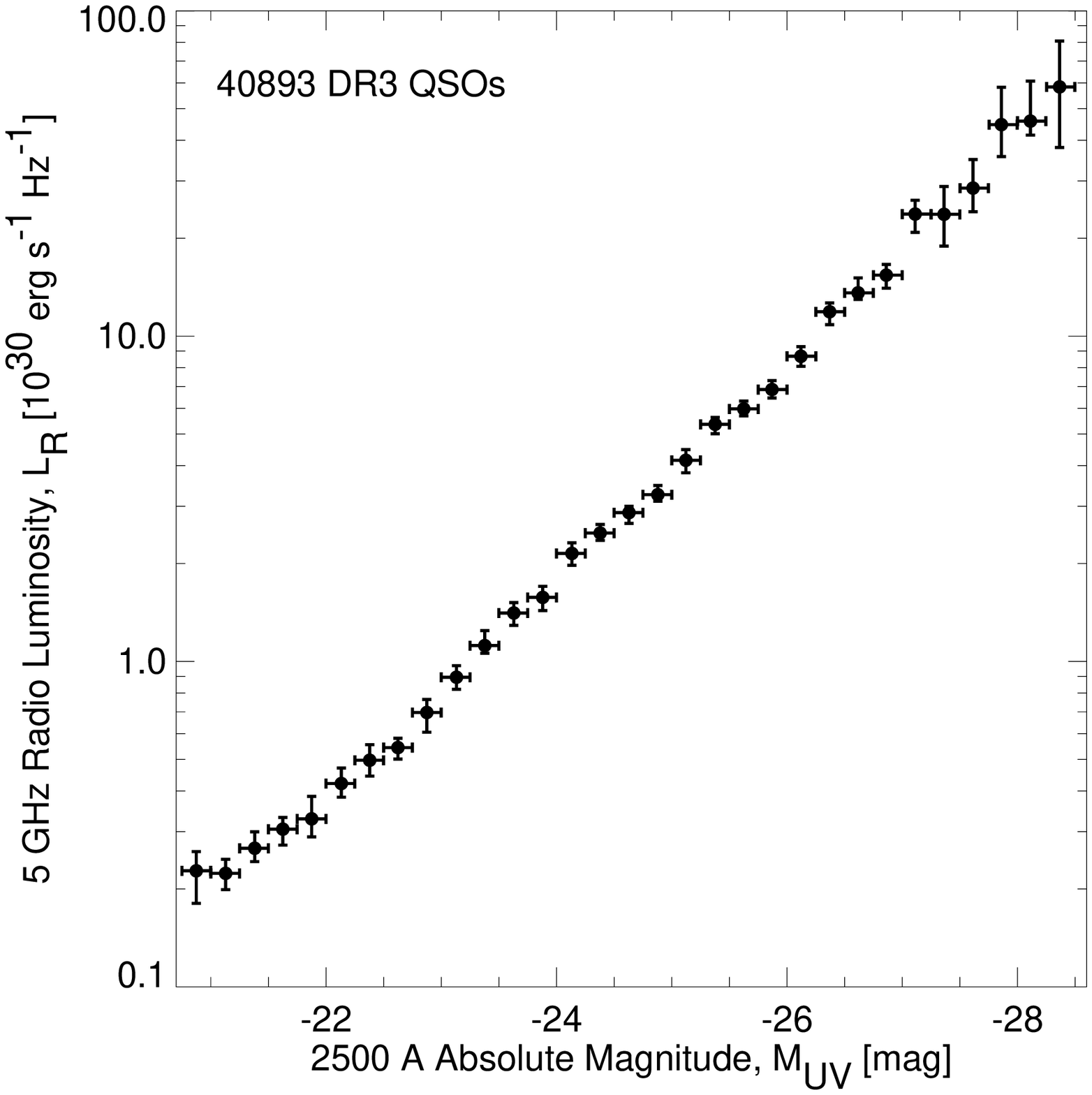}
\epsscale{1.0}
\caption{
The 5~GHz median radio luminosity \LR\ is very well correlated with
\MUV, the
absolute ultraviolet magnitude at 2500~\AA\ rest wavelength.
\label{fig-rlum_vs_absmag}
}
\end{figure}
We convert each FIRST cutout image to radio luminosity units using
the known quasar redshift and then stack those scaled images to
compute median radio luminosities.  Figure~\ref{fig-rlum_vs_absmag}
shows the very close correlation between \MUV\ and \LR, which
is well fitted by a power-law:
\begin{equation}
\log\,\LR = 0.54 - 0.339(\MUV + 25) \quad ,
\end{equation}
where $\LR$ is in units of $10^{30}\,\hbox{ergs}\,\hbox{s}^{-1}\hbox{Hz}^{-1}$.
If the radio luminosity were simply proportional to the optical
luminosity, the slope would be steeper ($-0.4$ instead
of $-0.339$).  This slope implies $\LR \sim L_{opt}^{0.85}$;
the radio loudness $R^*$ is a declining
function of optical luminosity, with the most luminous sources
($\MUV=-28.5$) having $R^*$ values that are lower by a factor
of 3 compared with the least luminous sources ($\MUV=-20$).
\begin{figure*}
\epsscale{0.8}
\plotone{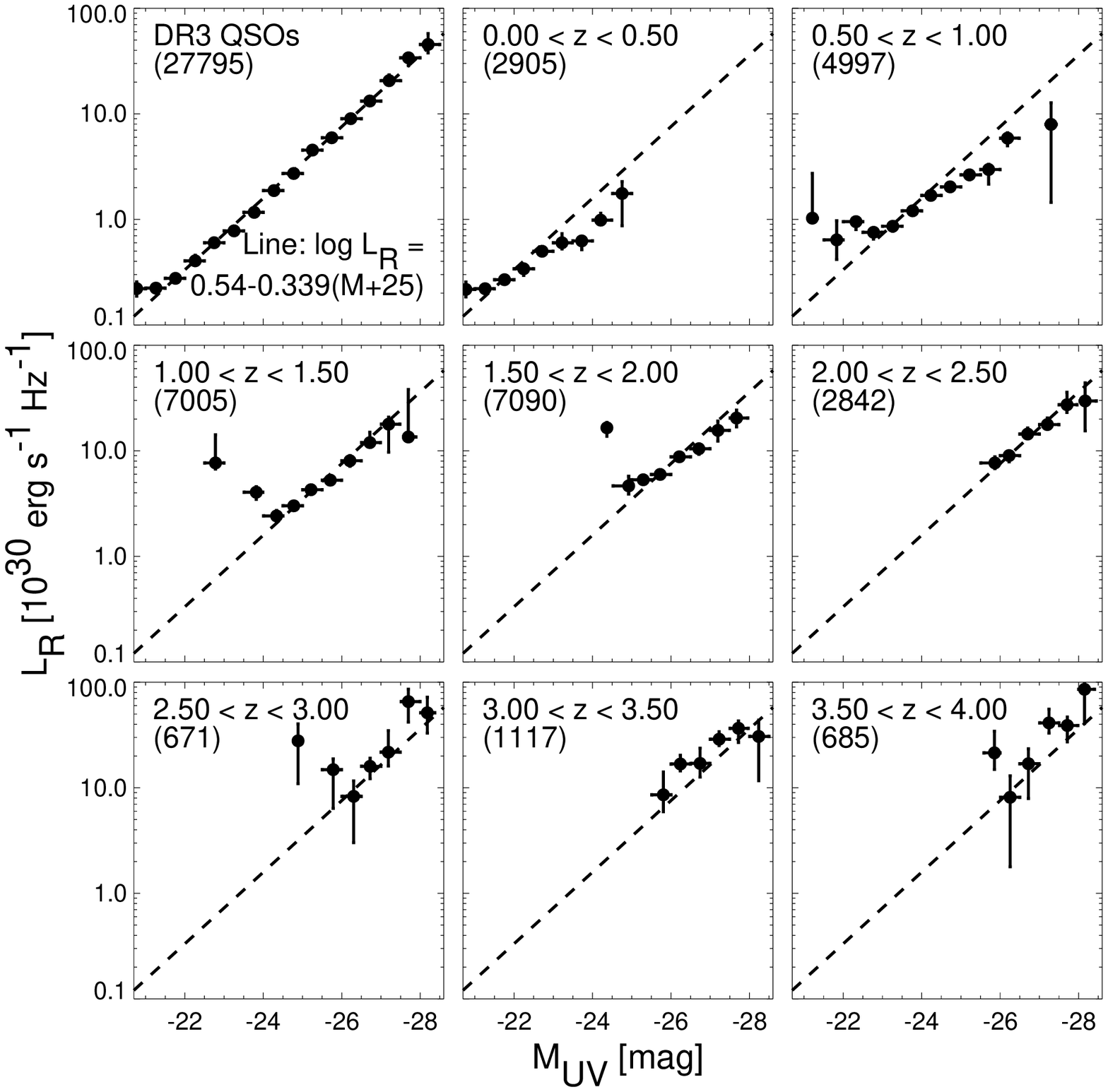}
\epsscale{1.0}
\caption{
The radio luminosity as a function of absolute magnitude with the
quasar sample divided into redshift intervals.  Only DR3 quasars
selected using the \prim\ or \hiz\ selection
criteria are included.  The upper left panel shows the combined
sample; the number in parentheses gives the number of sources in
each redshift interval.  Despite the strong absolute magnitude-redshift
correlation, which can make it difficult to separate dependencies
on the two variables, it is clear that the quasars in all the
redshift bins follow the same $\LR$ versus $\MUV$ relationship with at
most modest variations.
}
\label{fig-rlum_vs_absmag_z}
\end{figure*}
The absolute magnitude is strongly correlated with redshift, but if
the sample is divided into redshift intervals we find that a similar
$\LR$ versus $\MUV$ correlation applies at all redshifts
(Fig.~\ref{fig-rlum_vs_absmag_z}).  For this test we have restricted
the sample of DR3 quasars to those selected using the \prim\
or \hiz\ targeting criteria (Schneider et al.\ 2005).  The
\prim\ selection used $ugri$ colors to identify quasar
candidates at $z<3$ with magnitudes $i<19.1$; it includes 25,511
objects in regions covered by FIRST.  The \hiz\ criterion
used $griz$ colors to identify candidates to fainter levels ($i<20.2$) and at
redshifts greater than 3; our sample includes 2,412 such objects.
The bulk of the remaining DR3 quasars were selected using various
serendipity criteria.  We exclude them here because objects so selected
have unusual radio properties (as discussed further below in
\S\ref{section-redshift_color}.)

In order to remove this strong radio-optical correlation, we scale
the radio properties to the reference absolute magnitude $\MUV =
-25$.  This is accomplished simply by multiplying the FIRST cutout
by an $\MUV$-dependent factor:
\begin{equation}
\log\,S_M = 0.339(\MUV+25) + \log\,S \quad,
\end{equation}
where $S$ is the original radio flux density.  The adjustment to
the radio luminosity, \LRM, is similar, and the adjusted radio-loudness
ratio is:
\begin{equation}
\log\,\RM = -0.061(\MUV+25) + \log\,R^* \quad.
\end{equation}
In all cases the $M$ subscript indicates that the quantity has been
adjusted for the absolute magnitude dependence.

\subsection{Variation of radio properties with redshift and
color\label{section-redshift_color}}
\begin{figure*}
\plottwo{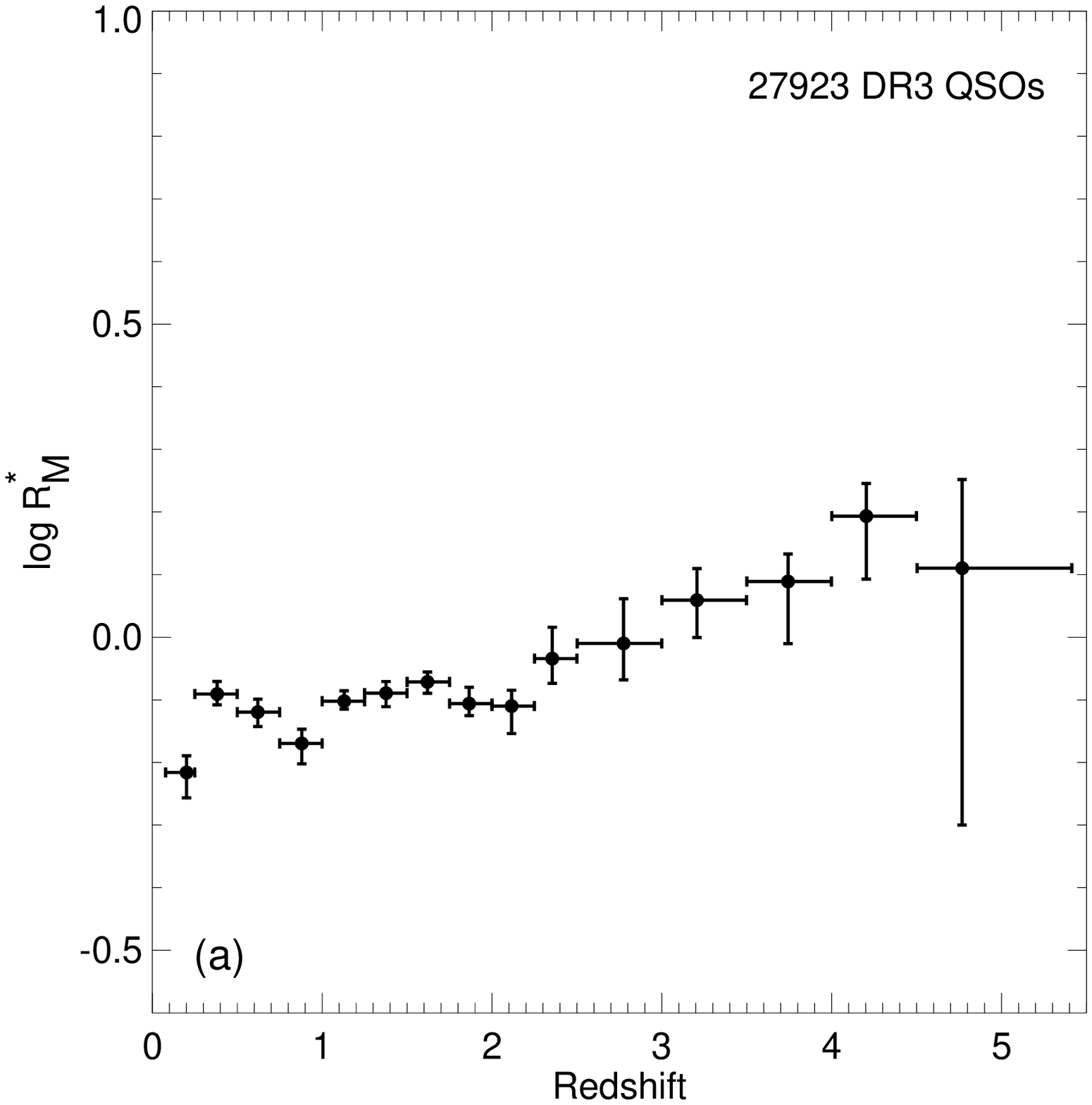}{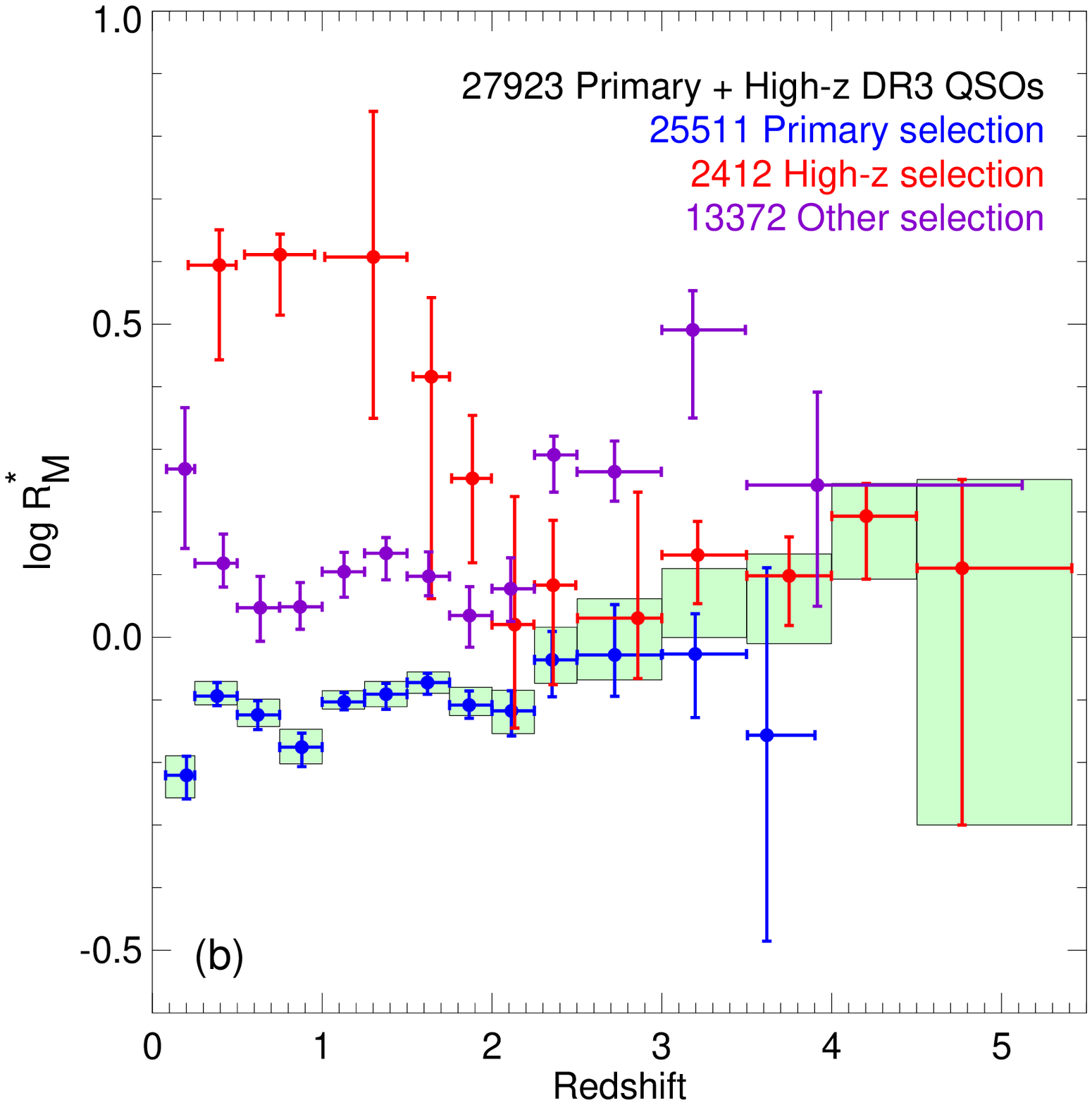}
\caption{
The dependence of the absolute-magnitude-adjusted radio loudness
\RM\ on redshift. (a) The distribution for the quasars selected
using the \prim\ and \hiz\ criteria is fairly flat, with \RM\ rising
by a factor of 2 between $z=0$ and $z=5$.
(b) When the sample is separated using the SDSS selection criteria
it is apparent that the radio properties vary greatly for different
SDSS samples.  The green boxes show the same combined distribution
from panel (a).
}
\label{fig-RM_vs_z}
\end{figure*}

\begin{figure*}
\epsscale{0.6} 
\plotone{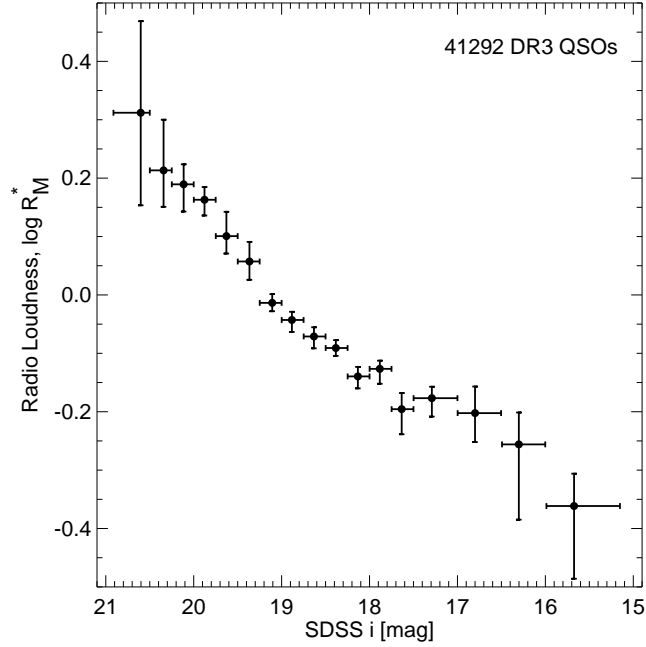}
\epsscale{1.0}
\caption{
The absolute magnitude-adjusted radio loudness \RM\ decreases toward
brighter $i$ magnitudes.
}
\label{fig-RM_vs_i}
\end{figure*}

\begin{figure*}
\epsscale{0.38} 
\plotone{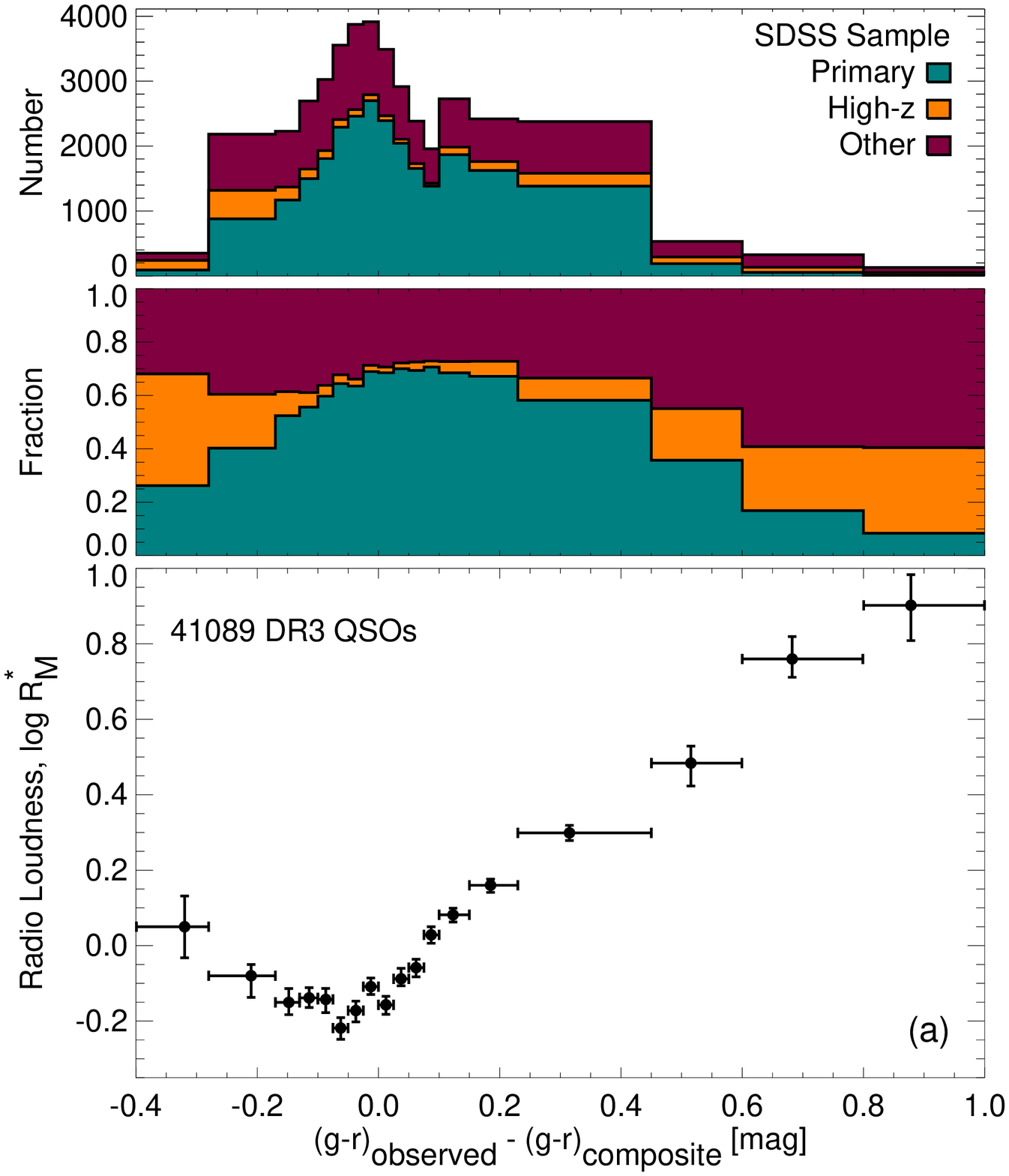}\hfil\plotone{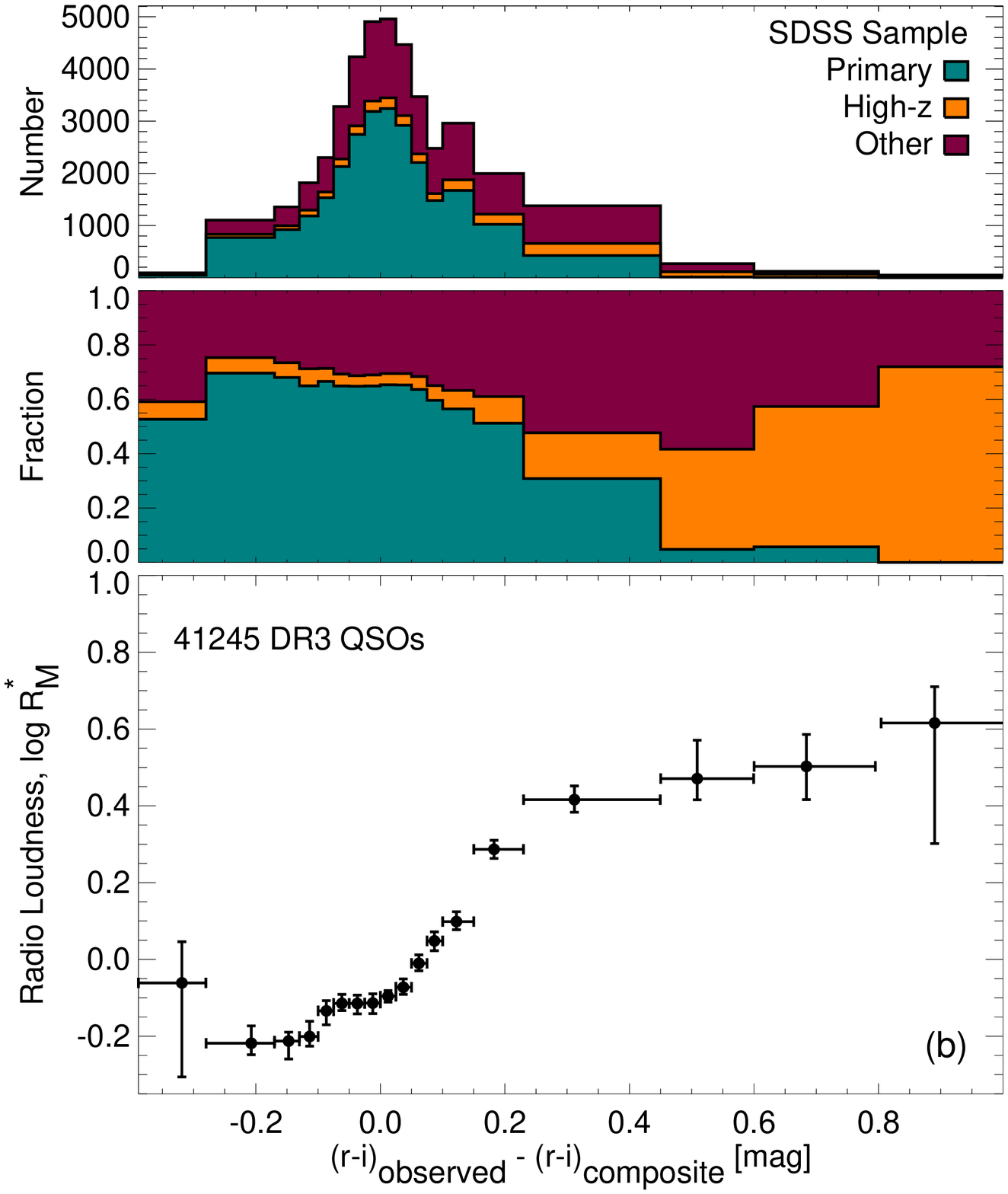}\hfil\plotone{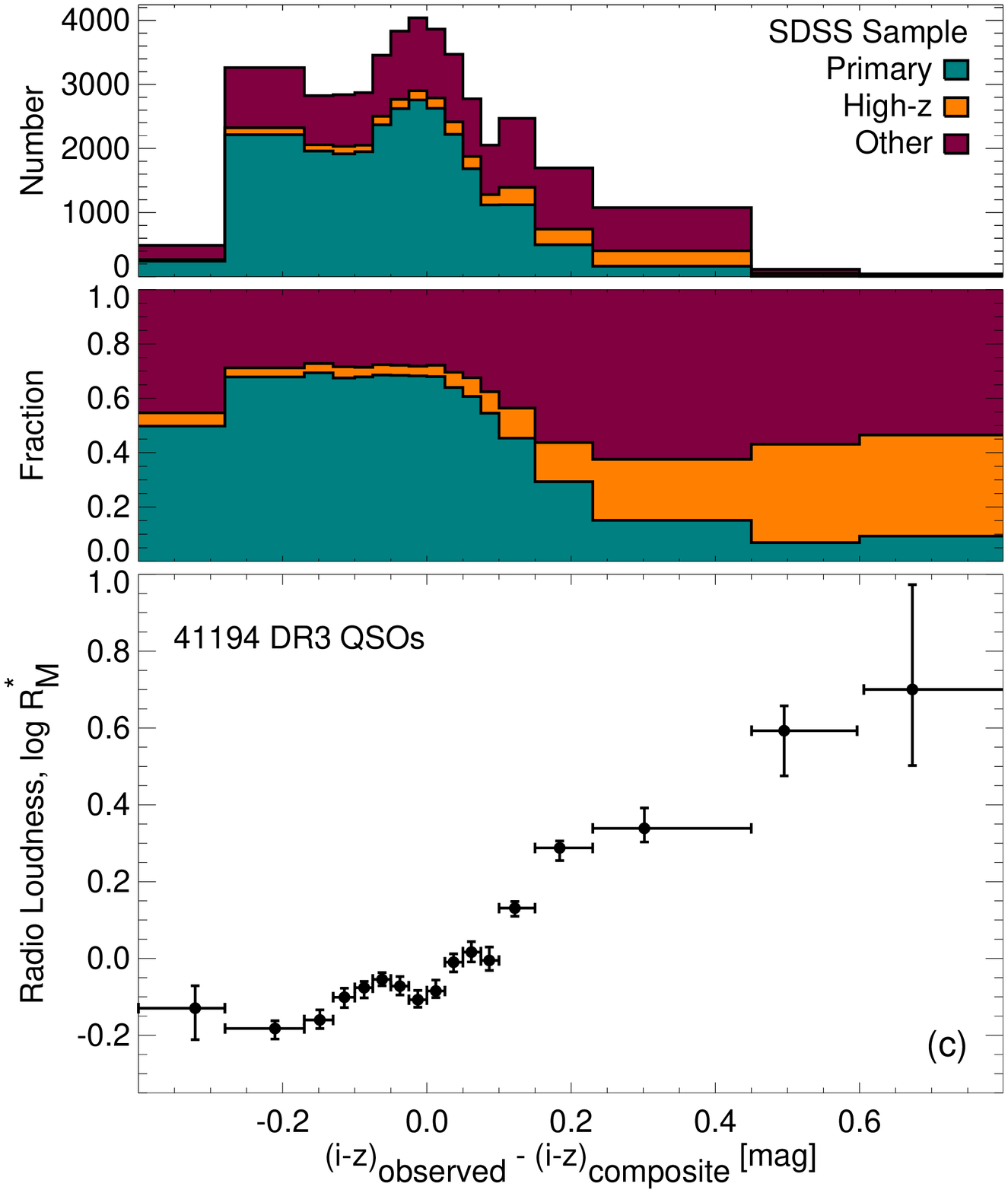}
\epsscale{1.0}
\caption{Dependence of the absolute magnitude-adjusted radio
loudness \RM\ on the SDSS optical colors.  The $x$-axis is
the difference between the observed SDSS colors and those predicted
for the standard SDSS composite quasar spectrum at the same redshift;
the nominal color is therefore zero.  The top plot shows the
histogram of the number of quasars in each color bin, with quasars
selected using different candidate criteria colored differently.
The middle plot shows the fraction of quasars selected by
the criteria in each bin; note that extreme colors are much less
likely to have been selected using the \prim\ criterion.
The bottom plot displays the mean radio loudness as a function
of color.
Quasars that are either redder or bluer than the composite
are much brighter in the radio.  The three panels show to
distribution for different SDSS colors ($g-r$, $r-i$, and $i-z$).
}
\label{fig-RM_vs_color}
\end{figure*}
Figure~\ref{fig-RM_vs_z}(a) shows the redshift dependence of the
radio loudness \RM\ after adjusing for the dependence on absolute
magnitude.  There is only modest evolution in this
quantity, with the radio-loudness being a factor of two higher at
$z=5$ than at $z=0$.  The radio properties of typical quasars
have changed little since the universe was one billion years old.

The picture changes, however, if we separate the SDSS DR3 sample
according to the criteria used to select the candidate quasars for
spectroscopic observations (Fig.~\ref{fig-RM_vs_z}b).  The quasars
selected using the \hiz\ criterion, which are redder and fainter
than the \prim\ candidates, are brighter in the radio.  This is
most noticeable at low redshifts ($z<2$), where the difference in
brightness is a factor of 4.  Even at high redshifts ($z>3$) a
slight difference persists; at least part of the slow rise in \RM\ toward higher
redshifts (Fig.~\ref{fig-RM_vs_z}a) appears to be created by the
transition in the SDSS sample from \prim-dominated selection for
$z<3$ to \hiz-dominated selection for $z>3$.

Quasars selected using other criteria (serendipity, ROSAT, FIRST,
stars, etc., as described in Schneider et al.\ 2005) are also
systematically radio louder.  One might be tempted to ascribe this
to the use of the FIRST catalog in selecting some of these candidates;
however, that introduces at most a very minor bias toward higher
\RM\ values.  Only 279 of the 13,372 sources selected using other
criteria are flagged in the DR3 catalog as FIRST sources, and
excluding them reduces the median radio loudness only slightly from
$\log\,\RM=0.11$ to $0.08$.  (The robustness of the median to the
presence of a rare admixture of bright sources is of course the reason
we choose to use it.)  Similarly, excluding ROSAT-selected sources ---
since radio emission is known to be more common among X-ray quasars
(e.g., Green et al.\ 1995) --- also leads to only a very small reduction
in \RM.  We conclude that there must be another explanation for
the different radio properties of the variously selected quasar
samples.

One possible contributor is the anti-correlation between radio
loudness and apparent magnitude (Fig.~\ref{fig-RM_vs_i}).  The
optically fainter sources are radio-louder, even after the \MUV\
adjustment.  Quasar candidates selected using the \prim\ criterion
are on average 1 magnitude brighter than those selected using other
criteria ($i=18.6$ versus 19.6).  But that accounts for a difference
in $\log\,\RM$ of only 0.15 and so does not explain the bulk of the
difference between the samples.

The most important underlying correlation that leads to differences
between the different SDSS samples is a strong dependence
of radio loudness on optical color (Fig.~\ref{fig-RM_vs_color}).
Since quasar colors
change systematically with redshift as various emission lines move
through the SDSS filters, we have subtracted the color of the SDSS
composite quasar spectrum (Vanden Berk et al.\ 2001) from
the observed colors.  That reduces
the scatter in colors and makes the expected color zero for a quasar
that resembles the composite.  We see a striking correlation:
quasars that are either bluer or redder than the standard color are
brighter in the radio, and substantially redder objects (with $g-r
> 0.8$ magnitudes) are brighter by a factor of $\sim10$ than quasars
with typical colors.
\begin{figure}
\epsscale{1.15} 
\plotone{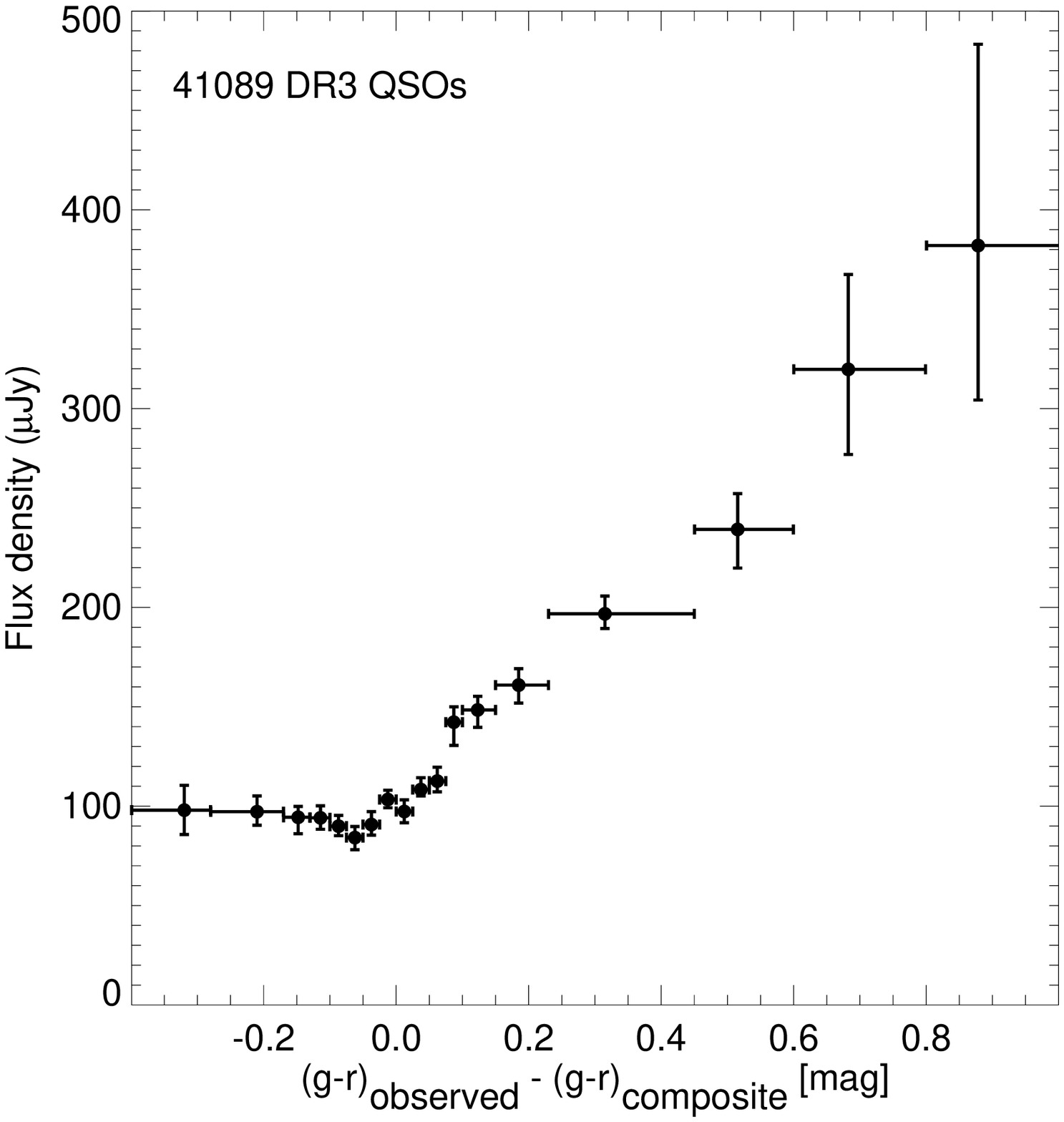}
\epsscale{1.0}
\caption{Variation in the median radio flux density as a function of
optical color.  The composite quasar color has been subtracted from
the observed SDSS $g-r$ color.  Redder sources have brighter fluxes,
with the reddest being $\sim2.5$ times the typical FIRST image
rms (145~\muJy).
}
\label{fig-flux_vs_color}
\end{figure}
The tendency of radio-loud quasars to have a larger scatter in their
optical colors has been noted before (Richards et al.\ 2001, Ivezi{\'c}
et al.\ 2002), although it has never been so clearly seen as in
this analysis.  Not only are the red quasars radio-louder, but their
median flux densities are also far higher (Fig.~\ref{fig-flux_vs_color}).
The increase in \RM\ for red objects is due primarily to brighter
radio fluxes, not to fainter optical magnitudes (which might also
be expected if the reddening is due to dust extinction.) Of the
factor of 10 variation seen in \RM, a factor of 4 is attributable to brighter
radio flux densities and a factor of 2.5 to fainter optical fluxes.  Note that
the reddest sources have median radio flux densities of nearly
0.4~mJy, tantalizingly close to detection by the FIRST survey.

Figure~\ref{fig-RM_vs_color} also shows the distribution of colors
for quasars selected using the various SDSS candidate criteria.
Quasars selected using the \prim\ criterion are much more concentrated
toward the typical (zero) colors than are objects selected by either
the \hiz\ or other criteria.  This makes a significant contribution
to the radio-loudness differences between the various samples
(Fig.~\ref{fig-RM_vs_z}b).  The color differences between the \prim\
and \hiz\ samples, when folded through the correlation in
Figure~\ref{fig-RM_vs_color}, lead to a difference in $\log\,\RM$ of
$\sim0.6$ between the samples for low-redshift quasars ($z<1.5$).
That accounts for most of the difference between the samples
seen in Figure~\ref{fig-RM_vs_z}(b).

The radio-color correlation also accounts for the bump seen in the
flux density at $z=2.25$ (Fig.~\ref{fig-flux_vs_z}).  The efficiency
of the SDSS color selection for quasars declines sharply in the
redshift range $2.4<z<3$ because the locus of normal quasar colors
crosses the stellar locus in the SDSS color space (Richards et al.\
2001, 2002).  The SDSS DR3 catalog contains far fewer objects in
this redshift range than might be expected based on the sensitivity
of the survey.  Moreover, the quasars that are included in the
catalog are dominated by objects with unusual colors compared with
the composite spectrum, since such objects do not resemble stars and so
can be selected by the usual SDSS criteria.  The jump in the radio
flux over this redshift range is created by the selection of a
larger fraction of redder quasars that have brighter radio emission
than normal quasars.

The observed increase in radio emission for bluer quasars can
be understood in the context of the unified model for AGN 
(Urry \& Padovani 1995) as being due to a beamed blazar
component affecting both the optical and radio emission.
The brightening for redder sources could also be attributed
to red optical synchrotron emission, but might instead
be explained as an evolutionary effect where dusty quasars are
more likely to be low-level radio emitters.  This is
likely to be a very useful clue to further understanding
of the origins of radio emission in AGN.

\subsection{The radio-loudness dichotomy\label{section-dichotomy}}
\begin{figure*}
\epsscale{0.8} 
\plotone{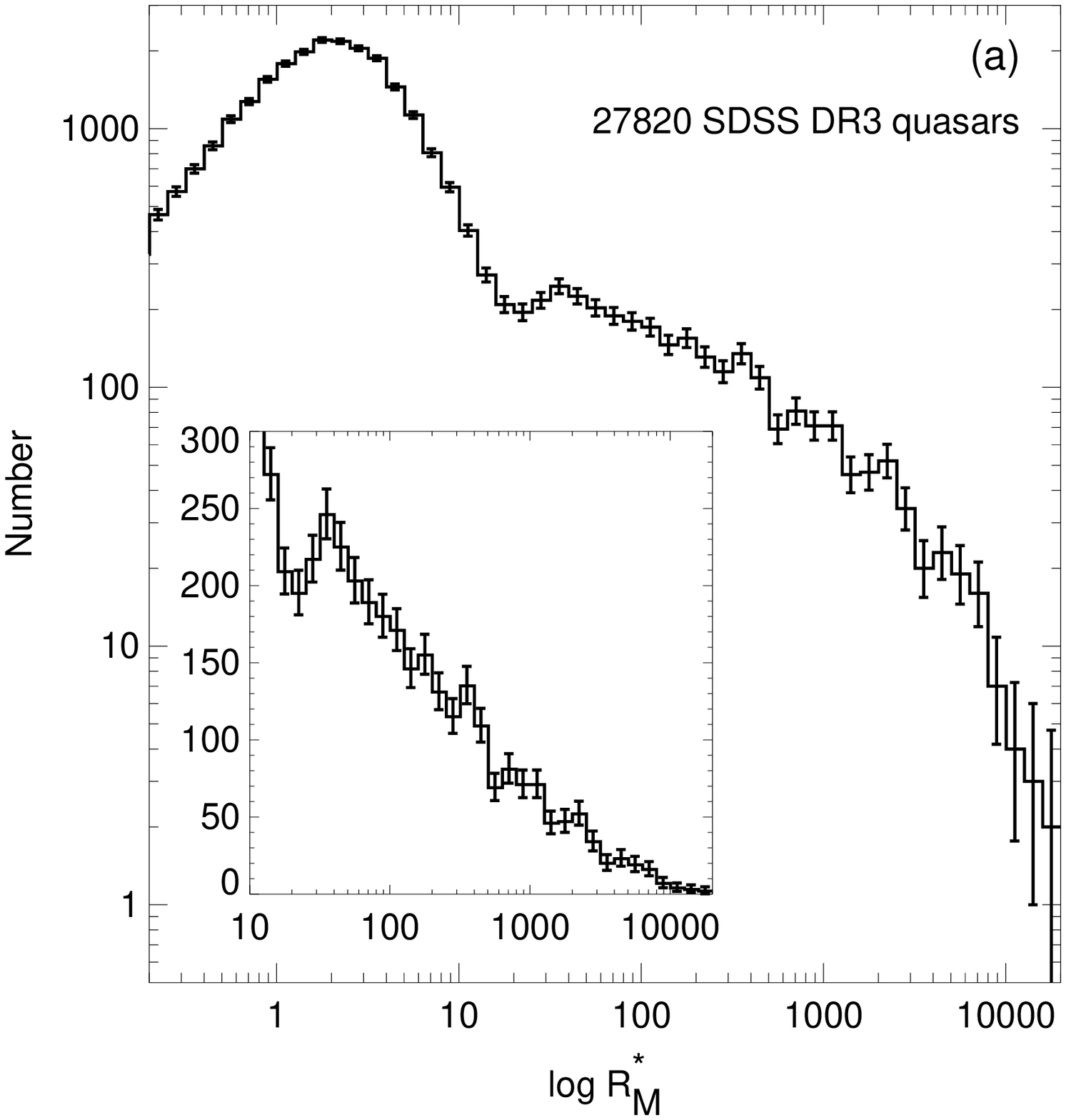}
\epsscale{0.38} 
\plotone{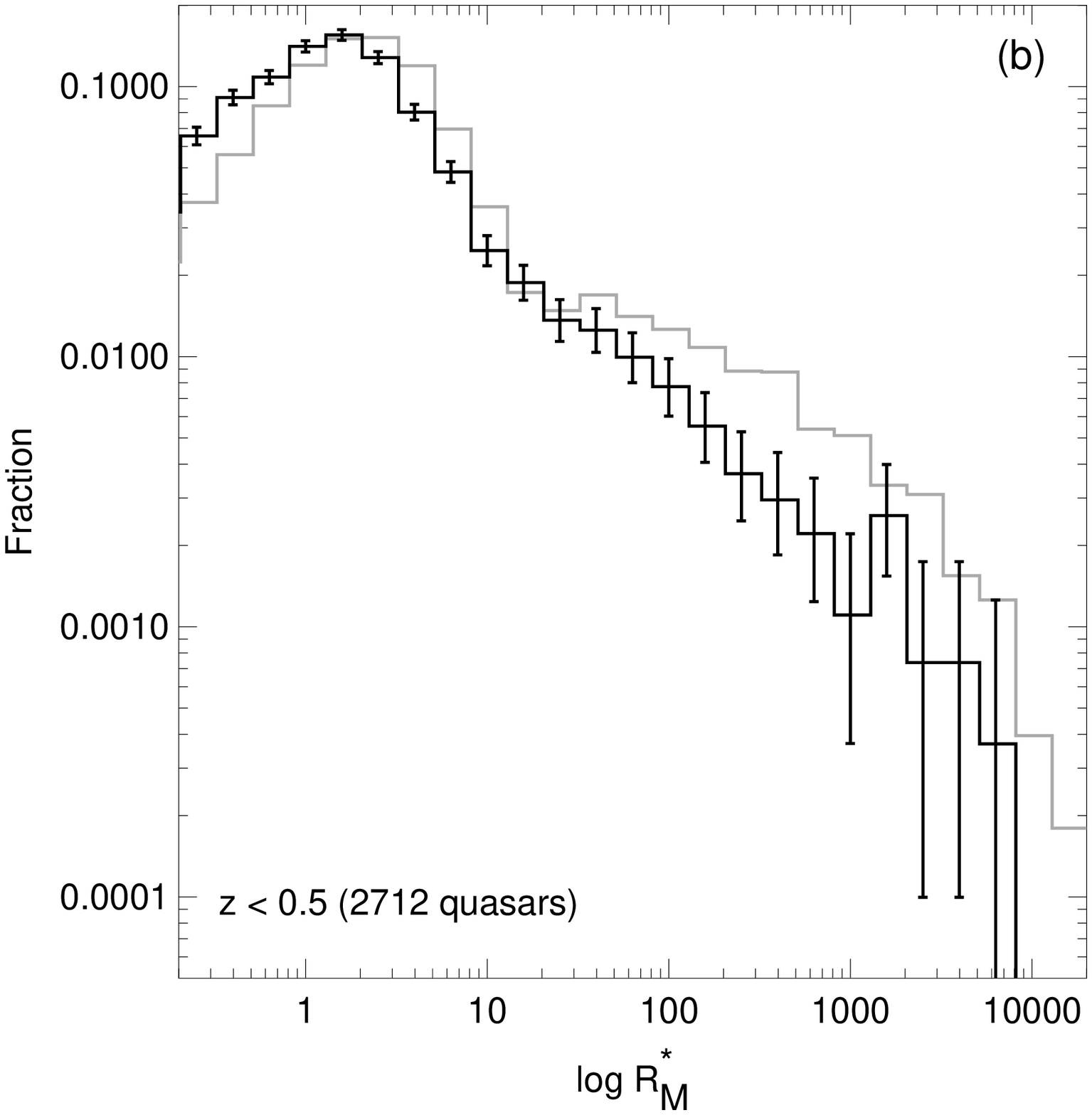}\plotone{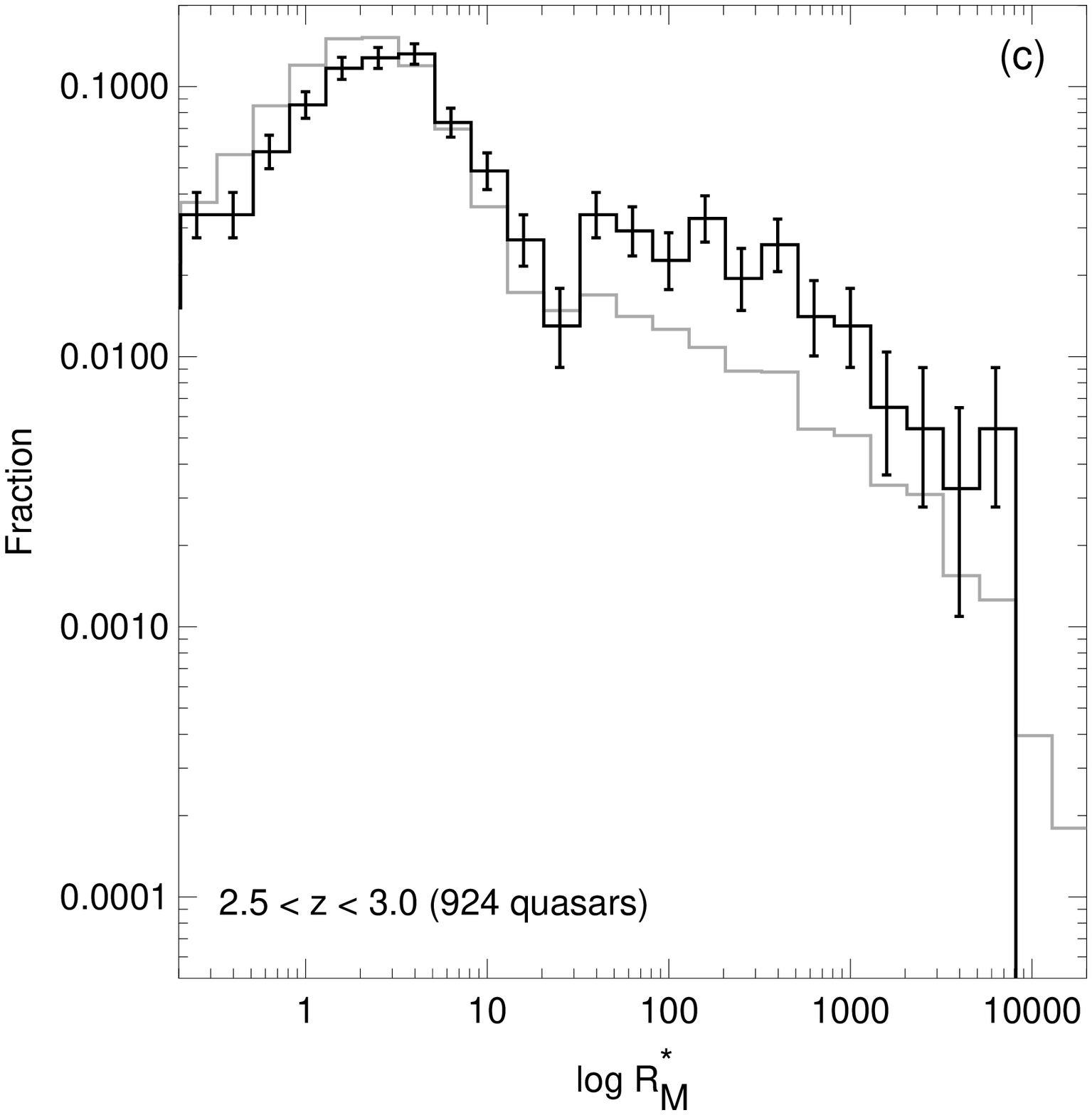}\plotone{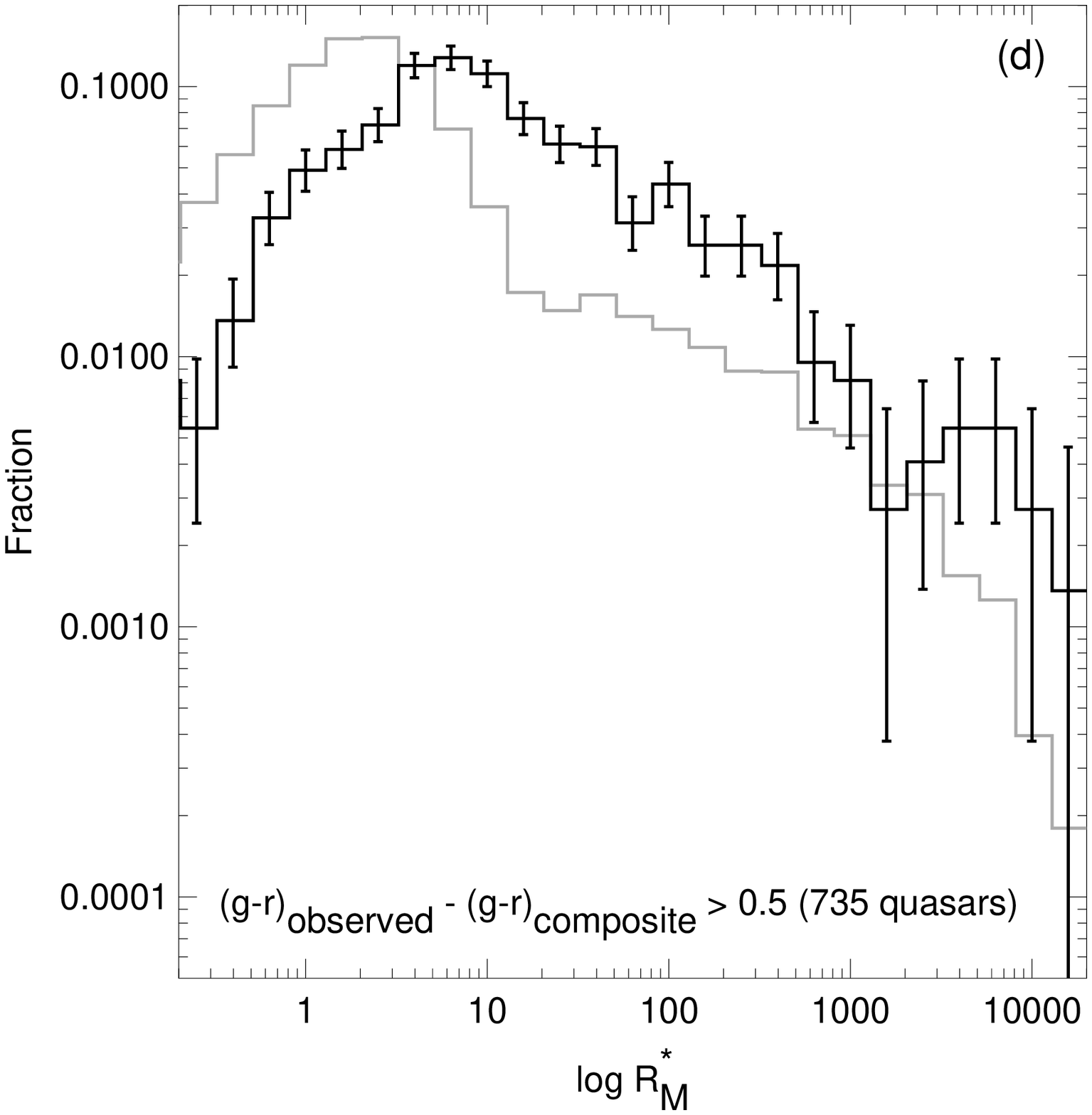}
\epsscale{1.0}
\caption{
The radio-loudness dichotomy as seen in the distribution of the
absolute magnitude-adjusted radio loudness parameter \RM.  (a) The
histogram for all SDSS DR3 quasars.  The inset shows an expanded
view with a linear $y$ scale.  There is a dip with an amplitude of
$\sim20\%$ separating radio-loud from radio-quiet objects, but the
dichotomy is not so clear in different parts of parameter space.
(b) Distribution for low-redshift quasars, normalized to show the
fraction in each bin. The gray line shows the distribution from
panel (a) for comparison.  (c) Distribution for quasars in the
redshift range $2.5<z<3$, where the SDSS selection effects are most
severe. (d) Distribution for red quasars.
}
\label{fig-R_hist}
\end{figure*}
Our absolute magnitude-adjusted radio loudness parameter \RM\ can
be used to re-examine the issue of whether the radio loudness
distribution is bimodal.  While all observers agree that there is
a highly non-Gaussian tail toward high $R^*$ values, White et al.\
(2000) and Cirasuolo et al.\ (2003a,b) did not see evidence for a
truly bimodal distribution with two peaks.  Ivezi{\'c} et al.\
(2002) did find a secondary peak, though their methodology was
questioned by Cirasuolo et al. Ivezi{\'c} et al.\ (2004) subsequently
applied the Cirasuolo et al.\ (2003b) approach to a large sample
of SDSS quasar candidates and claimed conclusive evidence for a
double-peaked distribution.

Figure~\ref{fig-R_hist} shows our distribution for the radio-loudness
parameter.  Note that the \RM\ parameter includes both redshift-dependent
$K$-corrections (as recommended by Ivezi{\'c} et al.\ 2004) and our
absolute-magnitude adjustment.  The overall distribution
(Fig.~\ref{fig-R_hist}a) clearly does show a secondary peak, although
the contrast in the valley between the peaks is considerably less
than the factor of two found by Ivezi{\'c} et al.\ (2004).  The peak is also
at a considerably lower $\log\,R^*$ value after the absolute magnitude adjustment
($\log\,\RM\sim1.55$ instead of $\log\,R^* \sim 1.9$).

An exploration of the dependence of the \RM\ distribution on other
parameters reveals a complex situation.  The radio-loud tail is
considerable weaker at low redshifts ($z<0.5$; Fig.~\ref{fig-R_hist}b)
but is especially strong in the intermediate redshift range ($2.5<z<3$;
Fig.~\ref{fig-R_hist}c) where the color-section effects discussed
above are dominant.  For red sources, the valley disappears altogether
with the resulting distribution being shifted by a factor of $\sim3$
toward higher radio-loudness (Fig.~\ref{fig-R_hist}c).  Our conclusion
is that there is indeed a double-peaked radio-loudness distribution for
SDSS DR3 quasars, but that the distribution varies dramatically
with redshift and color (and other parameters). The exact form of
the overall distribution is likely to have been sculpted by selection
effects, which must be modeled in detail before the relatively
modest 20\% dip between the radio-loud and radio-quiet quasars can
be interpreted.

\subsection{The radio emission of BAL quasars\label{section-bal}}

Historically, the strongest claim associating radio properties with
other quasar attributes was the absence of broad absorption lines
in radio-loud quasars (Stocke et al.\ 1992).  Becker et al.\ (2000)
showed that some radio-loud quasars have BALs, although they noted
that BALs still appeared to be absent from the most extreme radio-loud
objects. Recently, Trump et al.\ (2006) released a catalog containing
4784 BAL and near-BAL quasars from the SDSS DR3 release.  As with
quasars in general, most of these objects fall below the detection
threshold of FIRST, making them an ideal population for stacking
studies.  The FIRST survey covers 4292 of the cataloged BAL quasars.

Traditionally, BAL quasars have been divided into two primary subgroups,
high ionization (HiBALs) and low ionization (LoBALs).  The latter
are much rarer; in the SDSS/FIRST sample there are 3647 HiBALs and
only 645 LoBALs.\footnote{The Trump et al.\ (2006) catalog identifies
many sources as both HiBALs and LoBALs; we chose to make these
categories disjoint by labelling quasars as HiBALs only if they are
{\it not} LoBALs.} The HiBALs are identified primarily on the basis
of \ion{C}{4} absorption at 1550~\AA, while the LoBALs are identified
primarily from \ion{Mg}{2} absorption at 2800~\AA. As a
result, the known LoBALs tend to be at lower redshifts than the
known HiBALs.  HiBALs can only be identified at redshifts $z>1.7$,
while LoBALs can be recognized at redshifts as low as 0.5.
\begin{figure*}
\epsscale{1.15} 
\plottwo{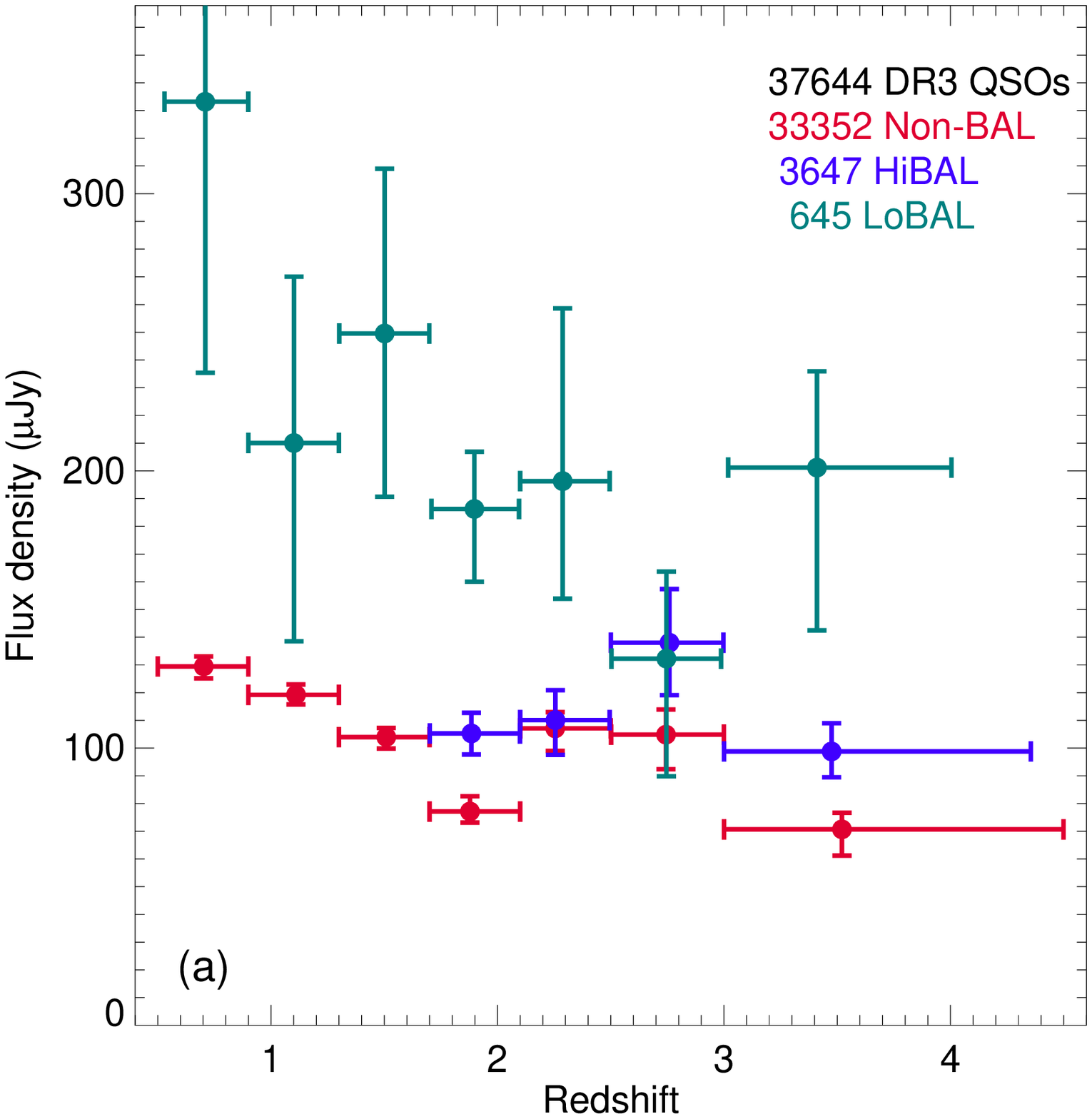}{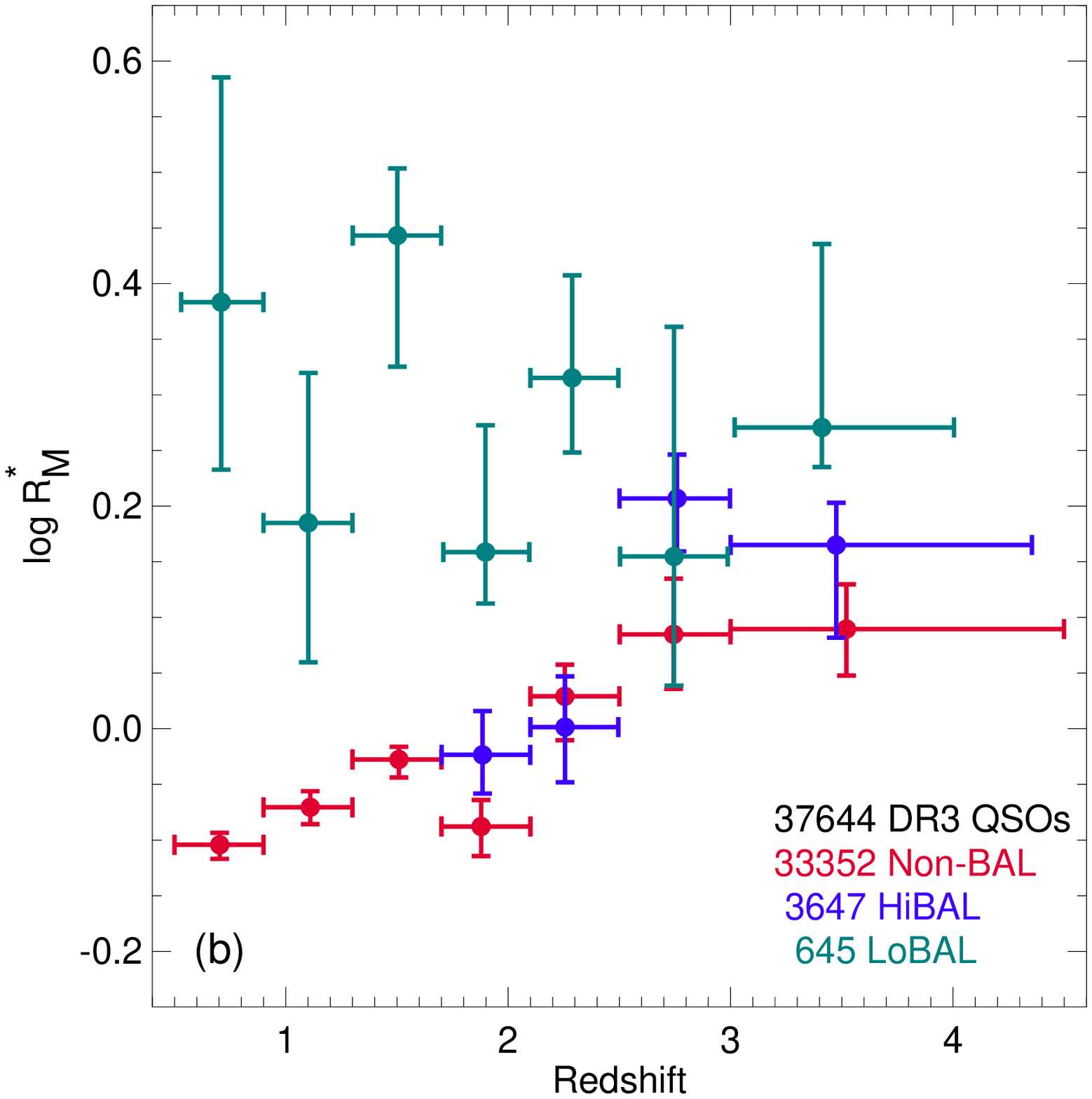}
\epsscale{1.0}
\caption{Median radio flux density (left) and absolute magnitude-adjusted
radio-loudness (right) for HiBAL, LoBAL and non-BAL quasars as a
function of redshift. Surprisingly, BAL quasars are brighter radio
sources than non-BALs, with the effect especially noticeable for
low-ionization BALs.
}
\label{fig-bal_vs_z}
\end{figure*}

\begin{figure*}
\epsscale{1.15} 
\plottwo{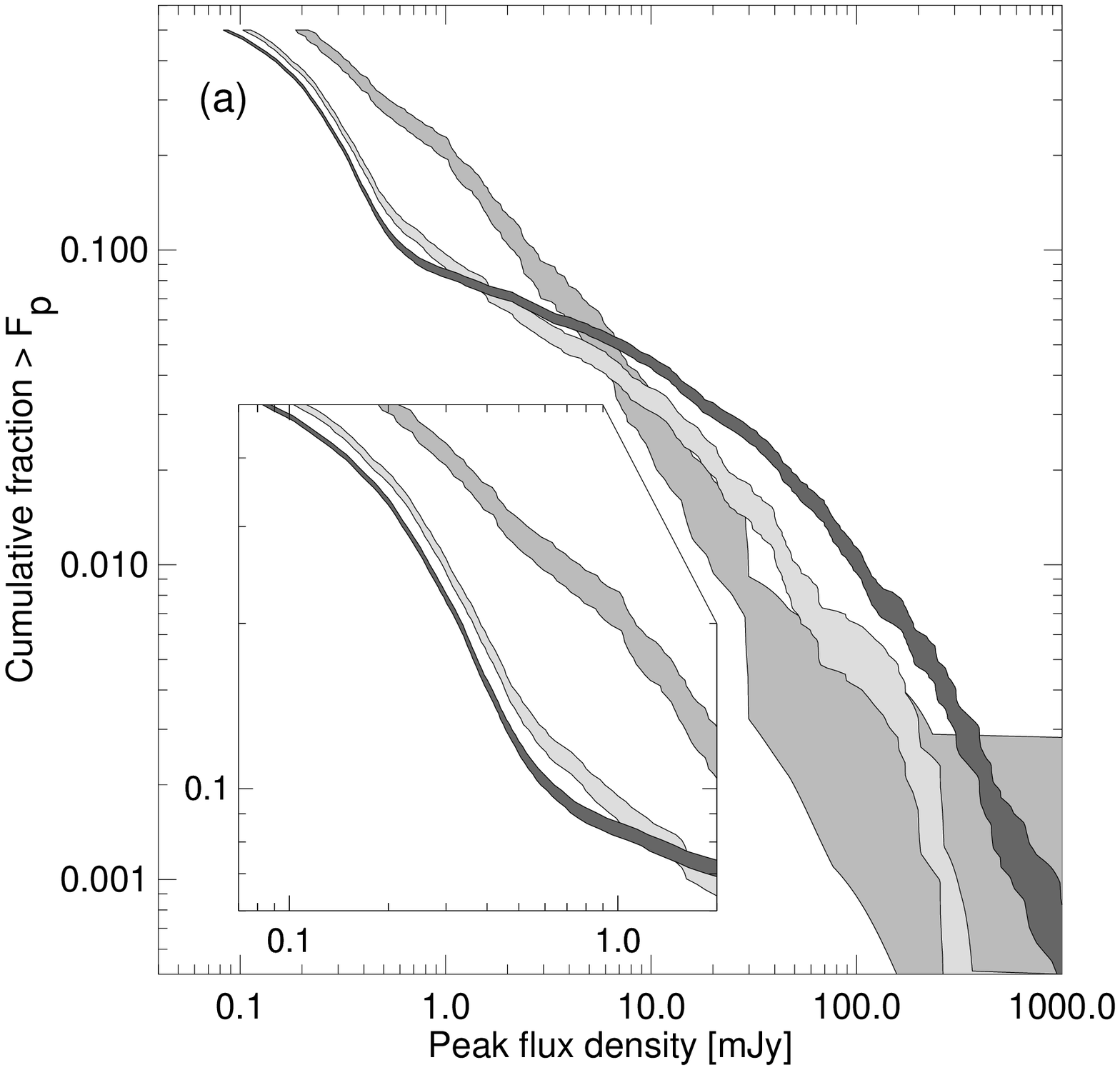}{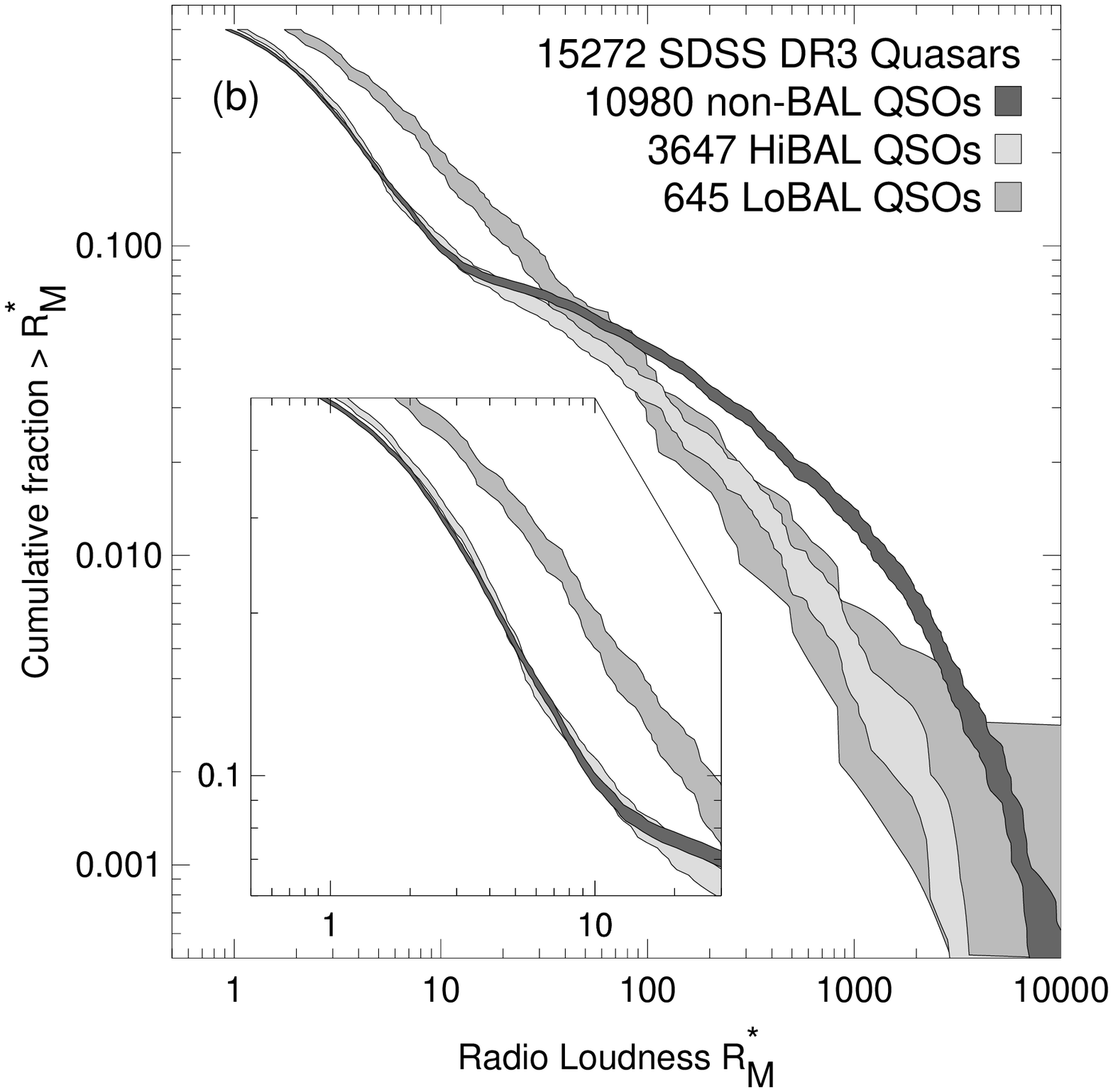}
\epsscale{1.0}
\caption{The cumulative distribution of HiBAL, LoBAL, and non-BAL
quasars as a function of flux density limit (left) and magnitude-adjusted
radio loudness \RM\ (right).  The shaded bands show $1\sigma$
uncertainties.  The bands converge at fraction 0.5 to the median
values derived from our stacking analysis; the inset shows an
expanded view of that region.  There is a deficit of BAL quasars
at bright fluxes, but there is an excess at fluxes fainter than
$\sim 1.5$~mJy for HiBALs and $\sim 5$~mJy for LoBALs.  The
radio-loudness distribution is similar for non-BALs and HiBALs in the
radio-intermediate region ($2<\RM<10$), though a small but significant
difference remains for radio-quiet ($\RM<2$) sources.  The LoBALs
are much more likely to be radio-intermediate sources than either
the HiBALs or non-BALs.
}
\label{fig-bal_cumplot}
\end{figure*}
In Figure~\ref{fig-bal_vs_z}(a), we show the median radio flux
density of HiBALs, LoBALs, and non-BALs as a function of redshift.
Interestingly, both classes of BAL quasars are brighter in the radio than
non-BALs.  The absolute-magnitude-adjusted radio loudness \RM\ shows
a similar effect (Fig.~\ref{fig-bal_vs_z}b), indicating that this
is not due to a difference in the distribution of \MUV\ for the
various classes.  The LoBALs are radio-louder by a factor $2.23\pm0.10$
(averaged over $0.5<z<4$) and the HiBALs by a 
factor $1.18\pm0.07$ ($1.7<z<4.3$).

That said, the comparison to the FIRST survey for this new large
sample of BAL quasars confirms the absence of extremely radio-loud BAL QSOs.
In Figure~\ref{fig-bal_cumplot}(a), we show the cumulative distribution
of BAL and non-BAL quasars as a function of radio flux density.  This
plot includes only non-BALs with $1.7<z<4$, since outside that
redshift range the absorption lines required for confident
identification of non-BALs do not fall in the SDSS spectrum window.
It is clear from the graph that while BALs are not found among the
brightest radio-emitting quasars, below 2~mJy they are systematically
brighter than non-BAL objects.

The disparity remains if we examine the radio-loudness parameter
instead of the flux density (Fig.~\ref{fig-bal_cumplot}b), although
the intermediate brightness HiBALs and non-BALs have \RM\ distributions
that are much more similar than their flux distributions.  A two-sided
Kolmogorov-Smirnov test shows the \RM\ distributions for HiBALs and
non-BALs in Figure~\ref{fig-bal_cumplot}(b) are different at the
$4\times10^{-5}$ level of significance. Separate tests for the
distribution with $\RM>14$ and the portion with $\RM<5.25$ show that
the bright and faint distributions are both discrepant
($2\times10^{-7}$ and $4\times10^{-6}$, respectively).  The differences
compared with the LoBAL distribution also highly significant
($<4\times10^{-3}$) despite the fact that the LoBAL sample is much smaller.
The \RM\ distributions for the radio-bright ($\RM>100$)
HiBAL and LoBAL quasars are statistically indistinguishible.
\begin{figure*}
\epsscale{0.60} 
\plotone{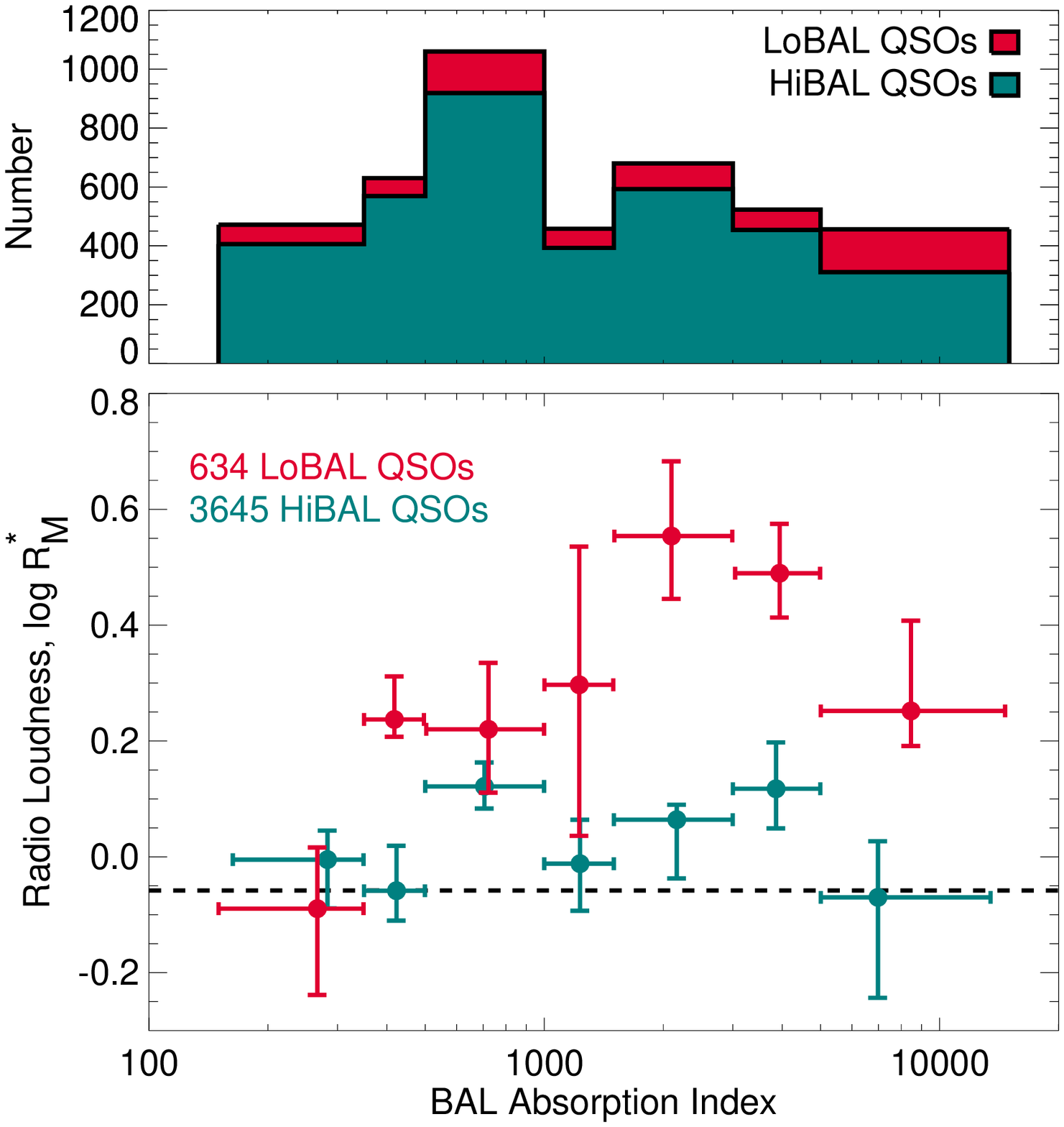}
\epsscale{1.0}
\caption{Median radio-loudness for HiBAL and LoBAL quasars as a
function of the BAL absorption index.  The top panel shows the
distribution of the absorption index values for the two types of
BALs.  The dashed line shows the radio loudness for the non-BAL
quasars (which by definition have absorption indices of zero.) There
is some indication of a drop in radio loudness for low absorption
indices, although the effect is not strong.
}
\label{fig-R_vs_absindex}
\end{figure*}
We have also examined the dependence of \RM\ on the BAL quasar
catalog's absorption index, which quantifies the strength and
extent of the broad absorption (Trump et al.\ 2006).  There
is a hint of a decline in \RM\ at the lowest absorption index
values, though the size of the effect is modest at best
(Fig.~\ref{fig-R_vs_absindex}).
\begin{figure*}
\epsscale{0.60} 
\plotone{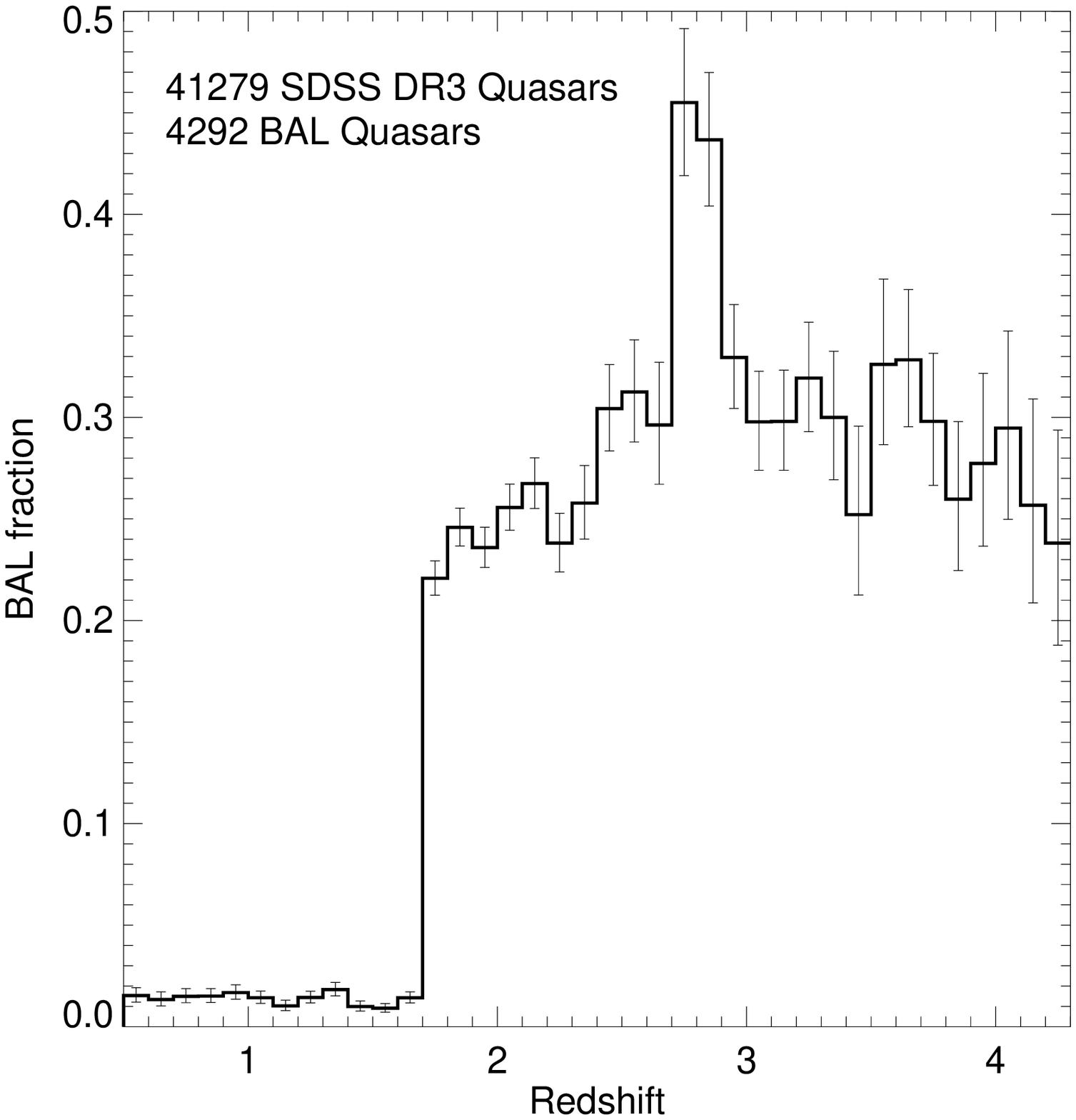}
\epsscale{1.0}
\caption{
Fraction of DR3 quasars that are BALs as function of redshift.
HiBALs can only be recognized when $z>1.7$, which accounts for the
large jump at that redshift; lower redshift quasars are all of the
rare LoBAL type.  The spike at $z=2.7$ is created by the inefficiency
of the SDSS quasar color selection where the quasar locus passes
close to the stellar locus.}
\label{fig-bal_frac}
\end{figure*}
We note in passing that the sample of BAL quasars in the SDSS DR3
catalog is strongly biased around $z\sim2.5$ by selection effects
that favor the discovery of objects with unusual colors.
Figure~\ref{fig-bal_frac} shows the fraction of BAL quasars as a
function of redshift.  For $2.7<z<2.9$, almost half the objects in
the DR3 catalog are BALs!  The effect here is similar to the bias
in favor of the discovery of red quasars in this same redshift
interval (discussed in \S\ref{section-redshift_color}).  If the
broad absorption lines change the quasar magnitude in even one of
the five SDSS filter bands, it is more likely to be recognized as
having colors inconsistent with the stellar locus.  This effect
has been previously discussed by Hewett \& Foltz (2003) and Reichard
et al.\ (2003).

We find the radio dominance of BAL over non-BAL quasars difficult
to reconcile with claims that BALs are largely the result of a
preferred orientation (e.g., Murray et al.\ 1995, Elvis 2000). In
fact, most of the arguments against the orientation model to date
have been based on radio observations. Zhou et al.\ (2006) argued
that radio variability observed in six BAL QSOs was strong evidence
that at least for some BALs, we were looking along the axis of the
radio jet.  Becker et al.\ (1997) made similar arguments on the
basis of the flat radio spectra observed for some BAL quasars. And
Gregg et al.\ (2006) used the FR2 radio morphology exhibited by
some BAL QSOs to argue against the need of a special orientation.

If the presence of BALs is determined by orientation, the greater
radio-loudness of BAL QSOs implies that we are looking closer to,
not farther from, the jet axis in quasars with BALs. We know of no
model that results in higher measured radio flux density from the
quasar core with greater angular distance from the jet direction.
Rather, relativistic beaming should enhance radio emission at small
angles to the quasar symmetry axis.  This is at odds with conventional
orientation models that require viewing angles closer to edge-on
for BAL quasars (e.g., Elvis 2000).

An alternative explanation is that quasars with low-level radio
emission in the nucleus have more BAL clouds and consequently are
more likely to show absorption.  BAL quasars may be in a special
evolutionary phase in which low-level radio emission confined near
the nucleus is accompanied by an excess of absorption clouds; when
the radio source breaks out to become truly radio-loud, it quickly
eliminates the clouds that are the source of broad absorption lines.
This is an evolutionary unification model (e.g., L{\'{\i}}pari \&
Terlevich 2006).

Richards et al.\ (2004) and Richards (2006) suggest that the sequence
from LoBALs to HiBALs to non-BALs may be the result of orientation
coupled with a gradual transition from a radiation-dominated,
accretion-disk wind to a MHD-dominated wind (Everett et al.\ 2001;
Everett 2005).  In this view, powerful radio-bright quasars have a
tightly collimated polar MHD wind in which BAL clouds have a small
covering factor, while radio-weak objects have more widely distributed
BAL clouds in an equatorial, radiatively accelerated wind coming
off the surface of the disk (see Fig.~5 of Richards et al.\ 2004).
This is a static unification model in which the 
magnetic field strength (with an associated radio source) acts as a
second parameter (in addition to orientation).

This picture does not necessarily predict enhanced low-level radio
emission from BAL quasars since it nominally produces monotonically
decreasing radio emission with increasing BAL strength.  However,
it could plausibly be modified to explain our observations.  A weak
radio source might elevate the disk wind above the equatorial plane,
increasing the BAL covering factor for lines of sight that are not
heavily absorbed by the disk, while a stronger radio source would
collimate the flow and reduce the BAL covering fraction.  Any model
that matches our observations will share this non-monotonic behavior
in order to produce BALs that are enhanced in the presence of
low-level radio emission but suppressed by bright radio emission.

\section{Summary and conclusions}

We have demonstrated that the average radio properties of sources
in the {\it FIRST} survey area can be derived even for populations
in which the individual members have flux densities an order of
magnitude or more below the typical field rms. Our median stacking
algorithm is robust, and, following the calibration of snapshot
bias derived herein, can be used to provide quantitative information
of average source flux densities into the nanoJansky regime. In our
application of this algorithm to the SDSS DR3 quasar catalog, we
establish the radio properties of quasars as a function of optical
luminosity, color and redshift.  The average radio luminosity
correlates very well with the optical luminosity, with $\LR \sim
L_{opt}^{0.85}$.  There is a very strong correlation between radio
loudness and color, with quasars having either bluer or redder
colors than the norm being brighter in the radio; objects 0.8
magnitudes redder than average in $g-r$ have radio loudness values 10 times 
higher than quasars with typical colors.  At faint flux densities, BAL
quasars actually have higher average radio luminosities and
radio-loudness parameters than non-BAL objects, a result inconsistent
with the conventional orientation hypothesis for BAL quasars.

The correlation between radio emission and color is an intriguing
clue to the nature of our FIRST-selected red quasars (Gregg et al.\
2002, White et al.\ 2003, Glikman et al.\ 2004a).  It suggests that
a wide-area radio survey only a factor of two deeper than the FIRST
survey might be capable of detecting the bulk of the reddened
population, which would shed light on the still controversial
question of what fraction of all quasars are highly reddened.

The success of stacking FIRST images to find the mean radio properties
of sub-threshold radio sources depends on the availability of large
target lists. As shown in this paper, the SDSS quasar sample is
ideal for these purposes. In fact the SDSS provides much more than
quasars. We are currently working on a study of the mean radio
properties of SDSS narrow-line AGN (deVries et al., in preparation).
In that paper, we explore the dependence of radio emission on
the strength of various emission lines, on associated star formation,
and on black hole mass.  We are also examining the radio properties
of star-forming galaxies taken from the SDSS spectroscopic survey
of galaxies (Becker et al., in preparation). There is no reason
that these studies must be limited to extragalactic samples. Other
astrophysical sources of weak radio emission include several classes
of stars. (We show in \S\ref{section-wd} that the average radio
flux density of white dwarfs is fainter than $\sim10~\muJy$.)
The number of stars with spectral classification is large
enough that it will be feasible to study the radio properties as
a function of spectral type.

\acknowledgments

Thanks to Gordon Richards for helpful comments on a draft of this paper.
The authors acknowledge the support of the National Science Foundation
under grants AST-00-98355 (RHB) and AST-00-98259 (DJH \& EG).  RHB's
work was also supported in part under the auspices of the US
Department of Energy by Lawrence Livermore National Laboratory under
contract W-7405-ENG-48.  RLW acknowledges the support of the Space
Telescope Science Institute, which is operated by the Association
of Universities for Research in Astronomy under NASA contract
NAS5-26555.

\end{document}